\documentclass[twocolumn]{aa} 

\usepackage[font=small,labelfont=bf,labelsep=period]{caption}
\usepackage{graphicx}
\usepackage{txfonts}
\usepackage{multirow}
\usepackage{subcaption}
\usepackage{romannum}
\def\kms {km\,s$^{-1}$}
\def\Rn2 {{\rm II}}

\def\co {$\rm ^{12}CO$}
\def\s {$\rm ^{13}CO$}
\def\c18o {$\rm C^{18}O$}
\def\xab {$ X_{\rm ^{13}CO}$/$X_{\rm C^{18}O}$}
\def\msun {\(M_\odot\)}
\usepackage{color}

\usepackage{threeparttable}

\begin{document}

        \title{Studies of the distinct regions due to CO selective dissociation in the Aquila molecular cloud}
        
        \author{Toktarkhan Komesh
                \inst{1,2,3}
                \and
                Willem Baan\inst{1,4}
                \and
                Jarken Esimbek\inst{1,5}
                \and
                Jianjun Zhou\inst{1,5}
                \and
                Dalei Li\inst{1,5}
                \and
                Gang Wu\inst{1,5}
                \and
                Yuxin He\inst{1,5}
                \and
                Zulfazli Rosli\inst{6}
                \and
                Margulan Ibraimov\inst{3}
        }
        
        \institute{Xinjiang Astronomical Observatory, Chinese Academy of Sciences, Urumqi 830011, PR China \\
                \email{komesh.t@outlook.com, baan@astron.nl, jarken@xao.ac.cn}
                \and
                University of the Chinese Academy of Sciences, Beijing 100080, PR China
                \and
                Department of Solid State Physics and Nonlinear Physics, Faculty of Physics and Technology, Al-Farabi Kazakh National University, Almaty, 050040, Kazakhstan 
                \and     
                Netherlands Institute for Radio Astronomy,  ASTRON, 7991 PD, Dwingeloo, The Netherlands 
                \and     
                Key Laboratory of Radio Astronomy, Chinese Academy of Sciences, Urumqi 830011, PR China
                \and     
                The Department of Physics, Faculty of Science, University of Malaya, 50603, Kuala Lumpur, Malaysia
        }
        
        \date{Submitted to A\&A \\
        Accepted 26 OCT 2020}
        
        % \abstract{}{}{}{}{} 
        % 5 {} token are mandatory
        
        \abstract
        {}
        {We investigate the role of selective dissociation in the process of star formation by comparing the physical parameters of protostellar-prestellar cores and the selected regions with the CO isotope distributions in photo-dissociation regions. We seek to understand whether there is a better connection between the evolutionary age of star forming regions and the effect of selective dissociation}
        {We used wide-field observations of the \co , \s , and \c18o \  (\,J\,=\,1\,-\,0) emission lines to study the ongoing star formation activity in the Aquila molecular region, and we used the 70\,$\mu$m and 250\,$\mu$m data to describe the heating of the surrounding material and as an indicator of the evolutionary age of the core.}
        {
                The protostellar-prestellar cores are found at locations with the highest \c18o \  column densities and their increasing evolutionary age coincides with an increasing 70$\mu$m/250$\mu$m\, emission ratio at their location. The evolutionary age of the cores may also follow from the \s\, versus \c18o \ abundance ratio, which decreases with increasing \c18o \  column densities.
                The original mass has been estimated for nine representative star formation regions and the original mass of the region correlates well with the integrated 70\,$\mu$m flux density. 
                Similarly, the \xab \ ratio, which provides the dissociation rate for these regions correlates  with the 70$\mu$m/250$\mu$m\, flux density ratio and reflects the evolutionary age of the star formation activity.
        }
        {}
        
        
        \maketitle
        %
        %-------------------------------------------------------------------

        \section{Introduction}
        The Aquila rift is part of a dark lane of cosmic dust flowing prominently through the central region of the galactic plane of the Milky Way, forming the Great Rift. 
        The two known sites of star formation in the Aquila Rift cloud complex, namely the Serpens South \citep{Bontemps2010} and the W40 H\,\Rn2 \ region \citep{Smith1985}, have become a hot spot for the study of star formation.
        Spitzer observations show that W\,40 and the embedded cluster Serpens South are near to one another on the sky, such that the Serpens South is seemingly part of the more evolved W\,40 region \citep{Gutermuth2008}.
        The distance to Serpens Main and W\,40 was recently measured to be 436\,pc, and the distance to Serpens South should be similar because the two sources are kinematically connected \citep{2017ApJ...834..143O}.
        \citet{2020ApJ...893...91S} estimated a mass of $\thicksim$ 1.4 $\times$ $10^5$\,\(M_\odot\) for the W\,40 giant molecular cloud, and  determined a distance of $\thicksim$ 474\,pc.
        
        Recently, \cite{2015A&A...584A..91K} presented the results of the Herschel Gould Belt survey (HGBS) observations in the Aquila molecular cloud AMC complex imaged with the SPIRE and PACS photometric cameras from 70\,$\mu$m \ to 500\,$\mu$m. 
        These latter authors identified a complete sample of starless dense cores and embedded (Class\,0\,-\,I) protostars in the AMC complex using the multi-scale, multi-wavelength source-extraction algorithm. 
        In this study, we use the 70 $\mu$m \ and 250 $\mu$m \ data to describe the heating of the surrounding material and as an indicator of the evolutionary age of the core.
        
        In this work, we encounter regions that are called photo-dissociation regions (PDRs) and regions where we suspect that selective dissociation also plays a role.
        We investigate the role of selective dissociation in the process of star formation by comparing the physical parameters of protostellar-prestellar cores and the distinct regions with the CO isotope distributions in PDRs. We seek to understand whether  or not there is a close relationship between the evolutionary age of star forming regions and the effect
        of selective dissociation.
        A detailed description of how selective dissociation works is provided in Section \ref{SD}.
        Breifly, as the star formation process progresses, far-ultraviolet photons start to regulate both the heating and the chemistry of the resulting PDR \citep{1985ApJ...291..722T}. 
        %However, there is no unified understanding of the PDR spatial structure yet. 
        Selective dissociation of CO is an expected phenomenon in PDRs resulting from the UV radiation  field \citep{1985ApJ...290..615G, 2004Sci...305.1763Y}. In this process, UV radiation selectively dissociates less-abundant
        CO isotopologues more effectively than more-abundant ones owing to different levels of self-shielding   \citep{1988ApJ...334..771V, 2007A&A...476..291L, 2014A&A...564A..68S}. 
        As several regions have been identified as distinct star formation regions in the $\rm H_2CO$ absorption map of the AMC \citep{2019ApJ...874..172K}, here we compare the \c18o \  distribution and the $\rm H_2CO$ absorption.
        
        In the present paper, we show the results of the \co , \s, \, and \c18o \  observations toward the W40 and Serpens South regions of the Aquila Rift.
        In Section 2, we present the details of the observations.
        The results and discussions  are described in Section 3. 
        Finally, we present our conclusions in Section 4.

        \section{Archival data}
        
        The $\rm ^{12}CO(1-0)$, $\rm ^{13}CO(1-0),$ and $\rm C^{18}O(1-0)$ data, simultaneously observed using the 13.7\,m 
        millimeter wave telescope of the Purple Mountain Observatory in Delingha, were taken from the Millimeter Wave Radio 
        Astronomy Database\footnote{http://www.radioast.nsdc.cn}. 
        The central position of the on-the-fly (OTF) observing pattern is $18^h30^m03^s$ -2\degr02\arcmin40\arcsec (J2000).
        The half-power beam width (HPBW) of the observing system is $\thicksim$ 50 \arcsec, the velocity resolution is 0.17 \kms \, 
        , and the system temperature ranged from 180 to 320\,K. 
        The $\rm ^{12}CO(1-0)$, $\rm ^{13}CO(1-0),$ and $\rm C^{18}O(1-0)$ data were smoothed to the same spatial resolution of 60 \arcsec  and the cell size is 30 \arcsec.
        The one-sigma noise levels of the \co, \s, \, and \c18o \  data are 0.5\,K, 0.35\,K, and 0.35\,K, respectively.
        Assuming a distance of 436\,pc for the Aquila complex, the spatial scale of the maps is 0.124\,pc\,arcmin$^{-1}$.
        
        Infrared 70\,$\mu$m \ and 250\,$\mu$m \ images \citep{2010A&A...518L.102A} were obtained from The Herschel Gould Belt Survey Archive\footnote{http://www.herschel.fr}.
        \begin{figure*}[hbt!]
                \centering
                \begin{subfigure}[b]{.47\linewidth}
                        \includegraphics[width=\linewidth]{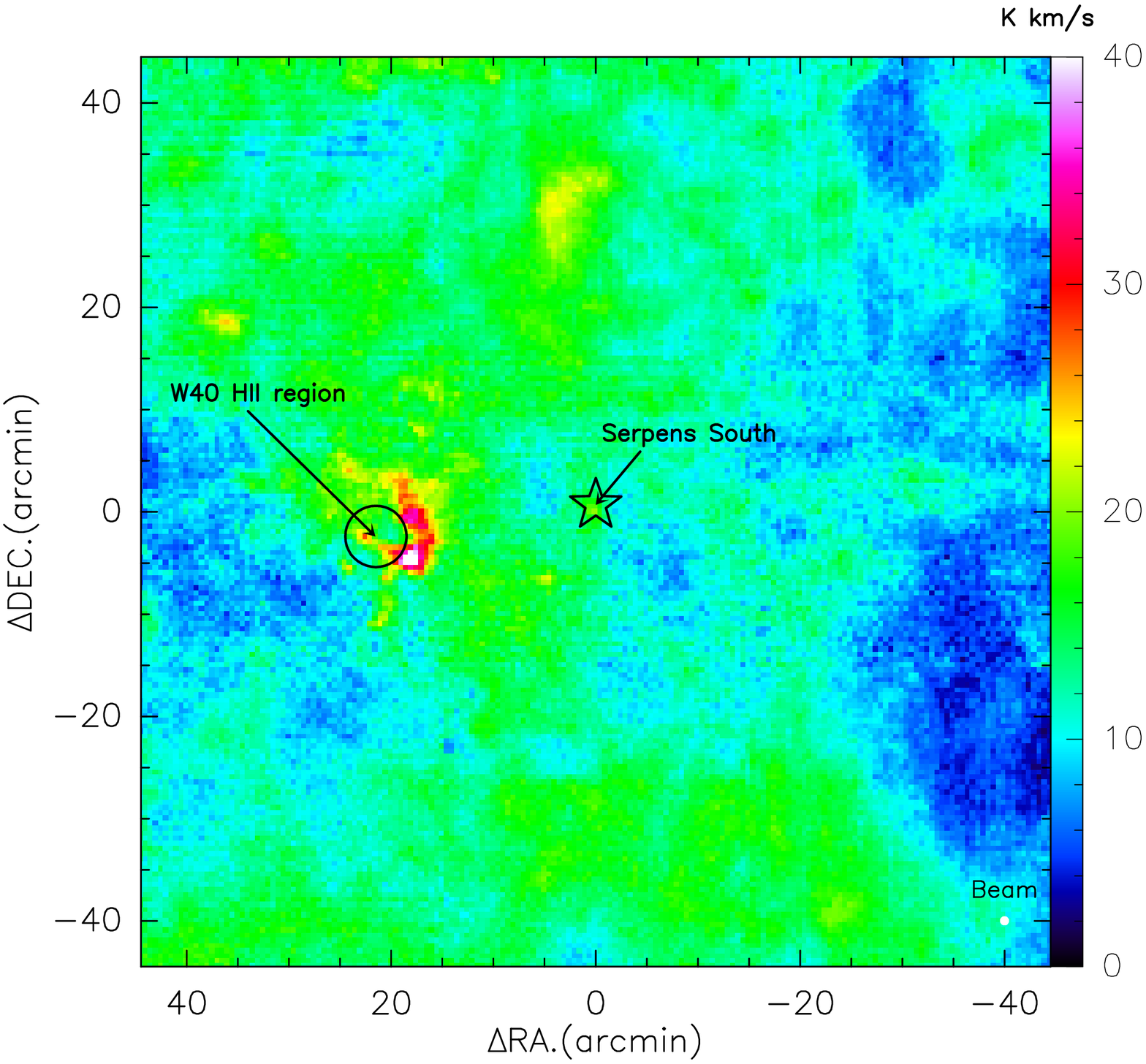}
                        \caption{}\label{fig:mouse}
                \end{subfigure}
                \begin{subfigure}[b]{.47\linewidth}
                        \includegraphics[width=\linewidth]{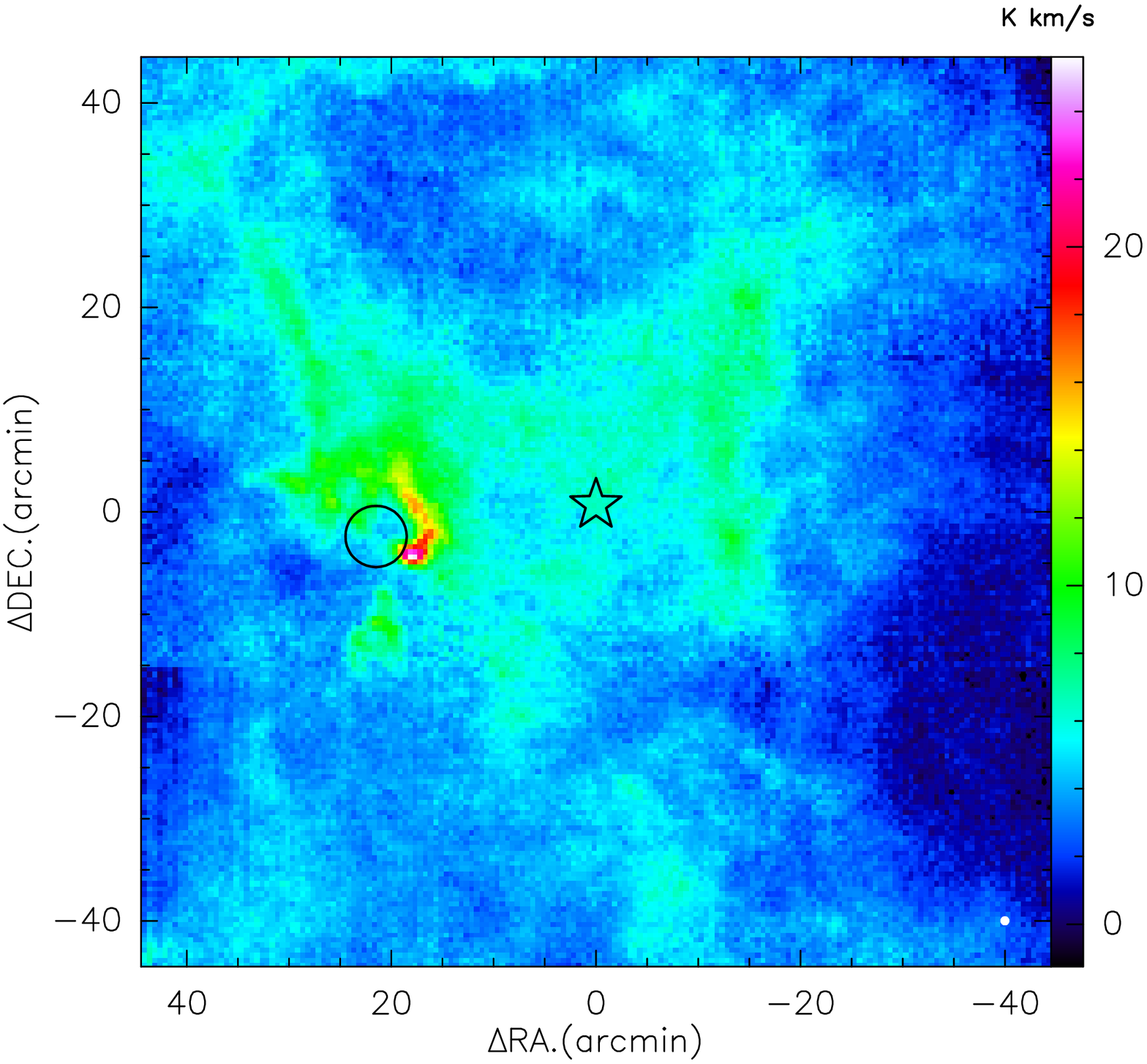}
                        \caption{}\label{fig:gull}
                \end{subfigure}
                
                \begin{subfigure}[b]{.47\linewidth}
                        \includegraphics[width=\linewidth]{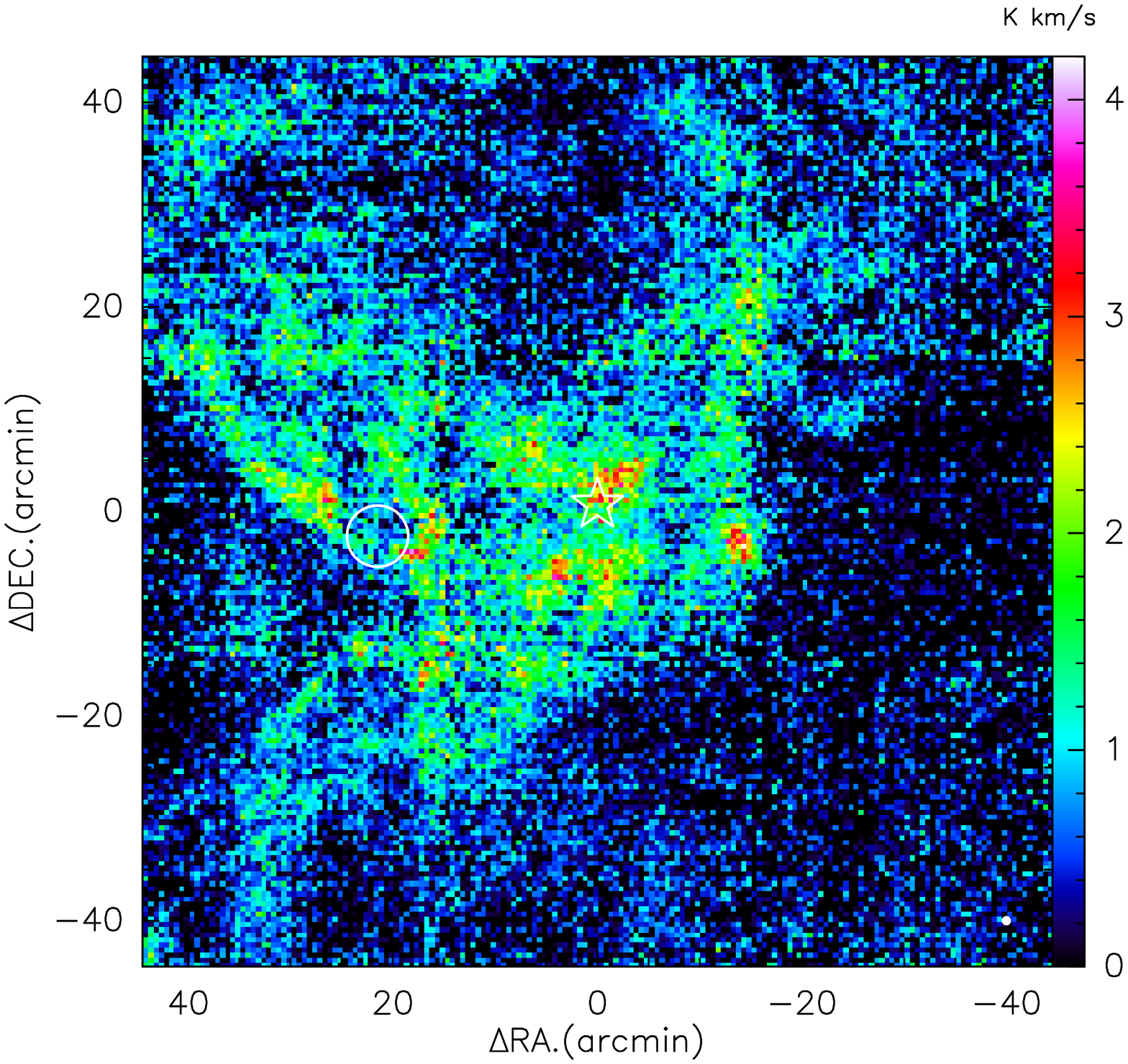}
                        \caption{}\label{fig:tiger}
                \end{subfigure}
                \begin{subfigure}[b]{.47\linewidth}
                        \includegraphics[width=\linewidth]{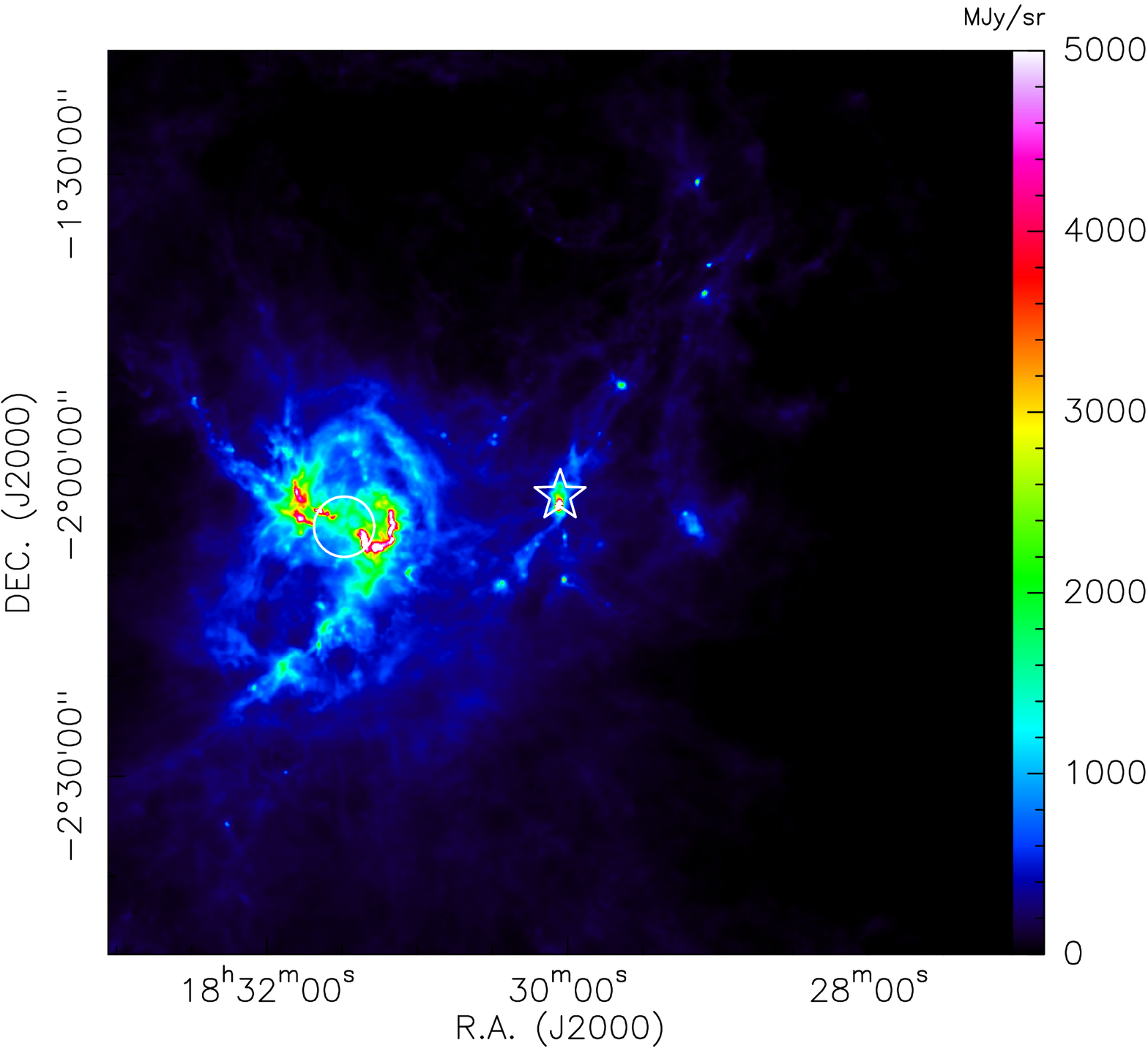}
                        \caption{}\label{fig:ti1ger}
                \end{subfigure}
                \caption{Aquila maps of the intensities of (a) $\rm ^{12}CO(1-0)$, (b) $\rm ^{13}CO(1-0)$, and (c) $\rm C^{18}O(1-0)$ integrated from  2 \kms to 12 \kms .  White filled circles in the bottom-right corner  of each panel illustrate the beam size of  60\arcsec.
                        (d) The Herschel 250\,$\mu$m \ image  (the beam size is 18.2\arcsec) taken from \cite{2010A&A...518L.102A}. The circle and star  mark the locations of the W40 H\,\Rn2 \ region (radius $\thicksim$ 3\arcmin) and the Serpens South, respectively. 
                }
                \label{fig:integ}
        \end{figure*}
        \begin{figure*}[hbt!]
                \centering
                \includegraphics[width=12cm]{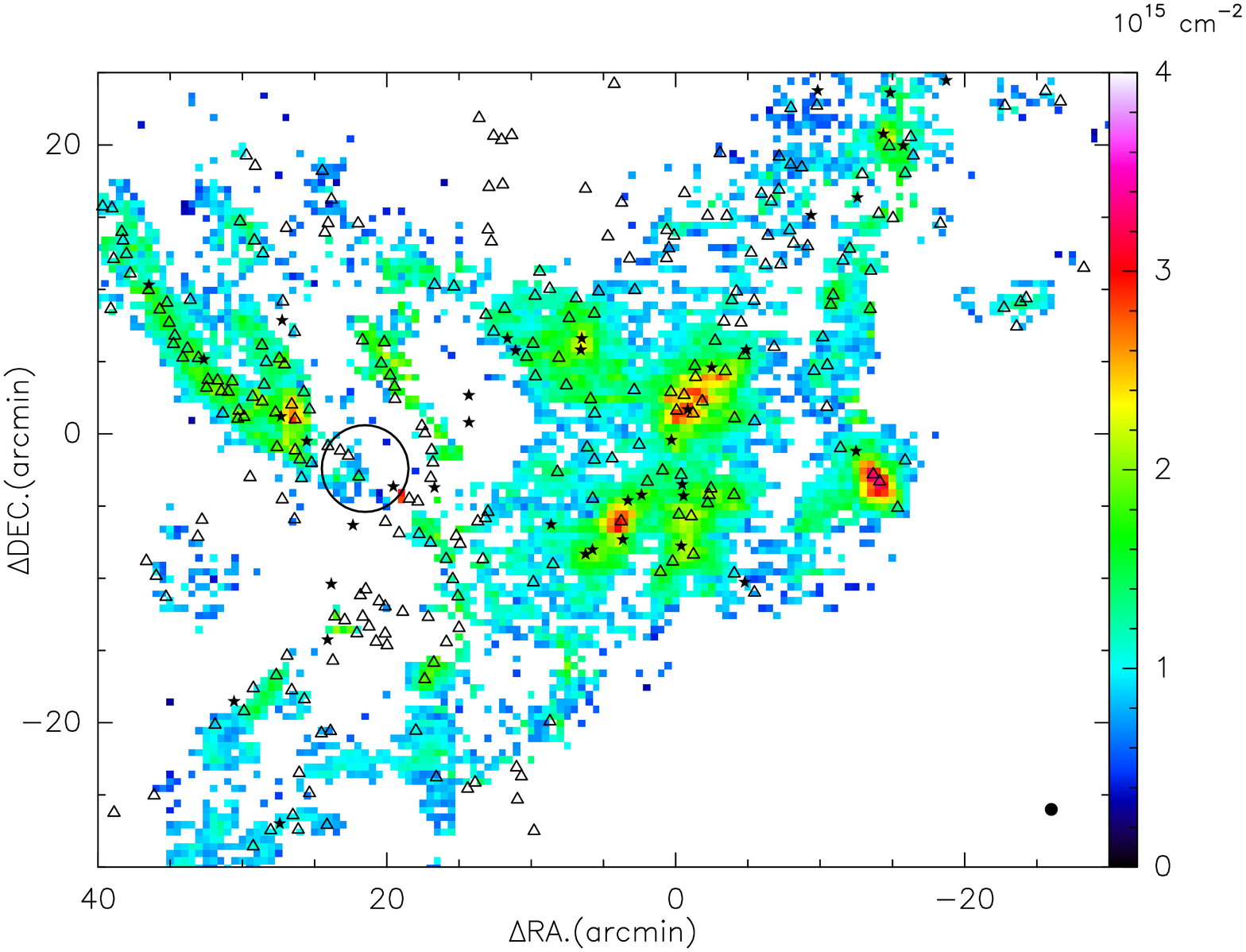}
                \caption{Map of the column density of $\rm C^{18}O(1-0)$. The black stars and triangles present the locations of protostellar and prestellar cores, respectively. The coordinates of the protostellar and prestellar cores were taken from \cite{2015A&A...584A..91K}. The black circle marks the location of the W40 H\,\Rn2 \ region (radius $\thicksim$ 3\arcmin) and the black filled circle in the bottom-right corner illustrates the beam size of  60\arcsec. The clipping level for the \s \ and \c18o \ data corresponds to a signal-to-noise ratio greater than five. Gaussian fitting to each pixel was done.}
                \label{fig:N}
        \end{figure*}
        
        \section{Results and Discussion}
        \subsection{$\rm ^{12}CO(1-0)$ Emission line}
        The integrated intensity map of $^{12}$CO(1-0) over a velocity range $2<V_{\rm LSR}<12$ \kms \ with signal-to-noise ratio greater than three is presented in Figure \ref{fig:integ}a. Enhanced core emission is present around W40 on the north and west sides, which could result from heating by the H\,\Rn2 \ region. There is weak emissions around Serpens South. The brightest position is at (RA,Dec) = ($18^h31^m15^s, -2\degr 08\arcmin 13\arcsec $).

        \subsection{ $\rm ^{13}CO(1-0)$ emission line}
        The integrated intensity map of $\rm  ^{13}CO(1-0)$ over a velocity range $2<V_{\rm LSR}<12$ \kms \  and a signal-to-noise ratio greater than three is presented in Figure \ref{fig:integ}b. Here, enhanced emission is also seen to the north and west of W40.  There are several elongated structures at Serpens South.

        \subsection{$\rm C^{18}O(1-0)$ emission line}
        The integrated intensity map of $\rm  C^{18}O(1-0)$ over a velocity range $2<V_{\rm LSR}<12$ \kms \ is presented in Figure \ref{fig:integ}c with its signal-to-noise ratio greater than three. Several integrated intensity concentrations are present in this map including both the W40 H\,\Rn2 \ region and Serpens South. The distribution of the $\rm C^{18}O(1-0)$ emission shows some similarity to the Herschel 250\,$\mu$m \  large-scale ($\thicksim$0.124\,pc) image presented in Figure \ref{fig:integ}d.  
        Both the \c18o \  and the Herschel data trace the same regions, which suggests that the \c18o \  and the Herschel 250\,$\mu$m \ emission arise from similar regions. 
        
        \begin{figure*}[hbt!]
                \centering
                \begin{subfigure}[b]{.335\linewidth}
                        \includegraphics[width=\linewidth]{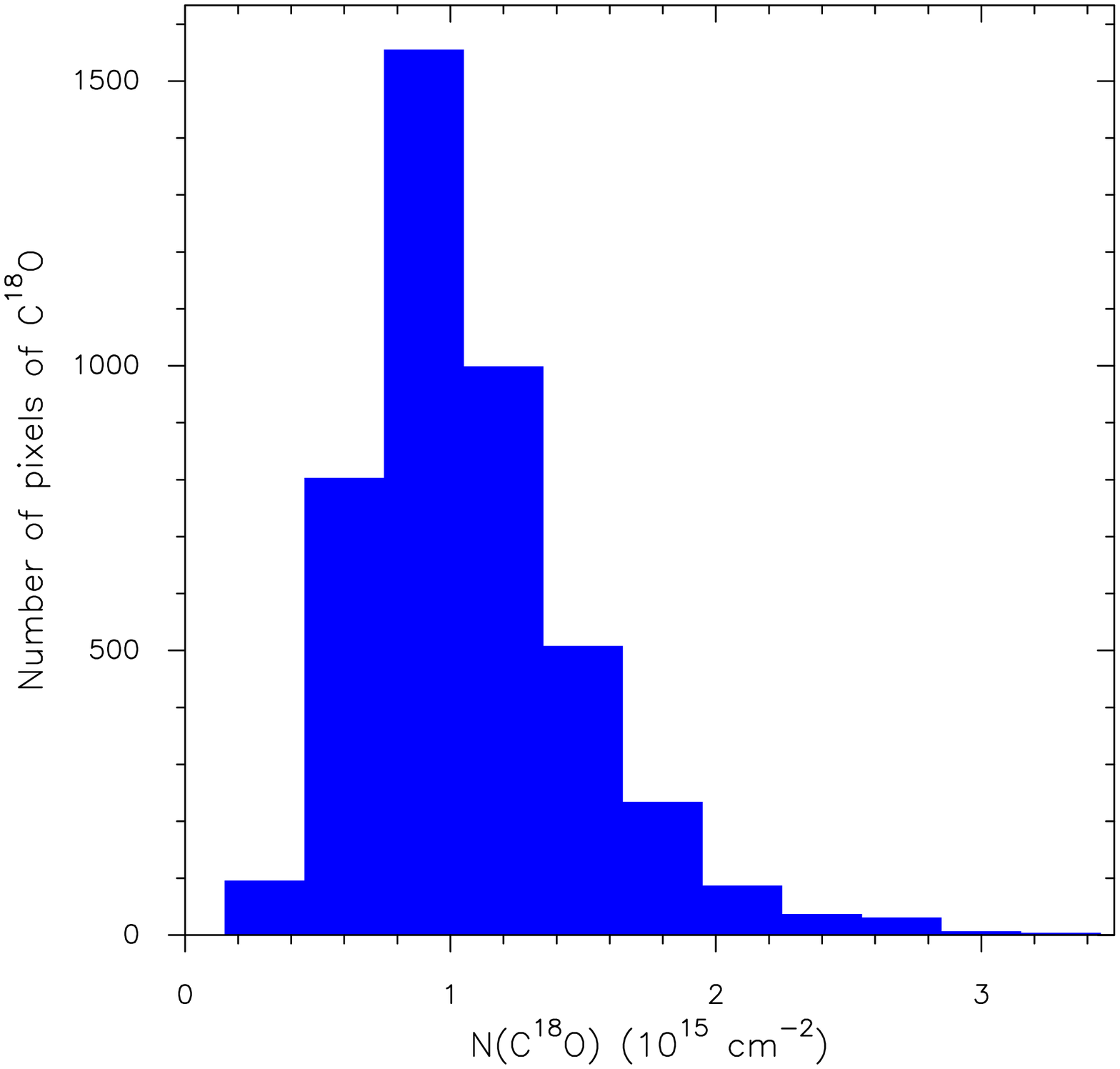}
                        \caption{}
                \end{subfigure}
                \begin{subfigure}[b]{.32\linewidth}
                        \includegraphics[width=\linewidth]{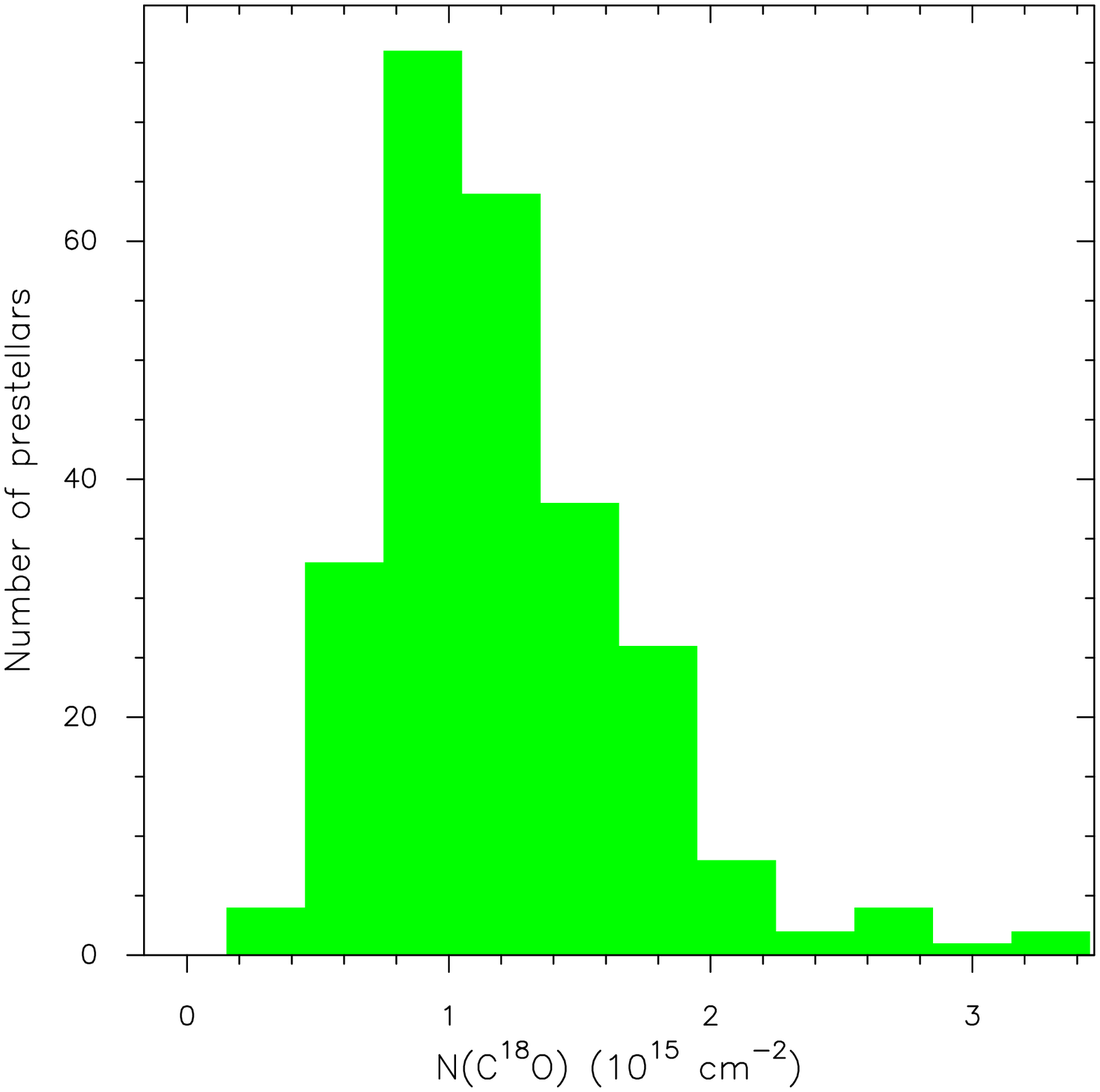}
                        \caption{}
                \end{subfigure}
                \begin{subfigure}[b]{.32\linewidth}
                        \includegraphics[width=\linewidth]{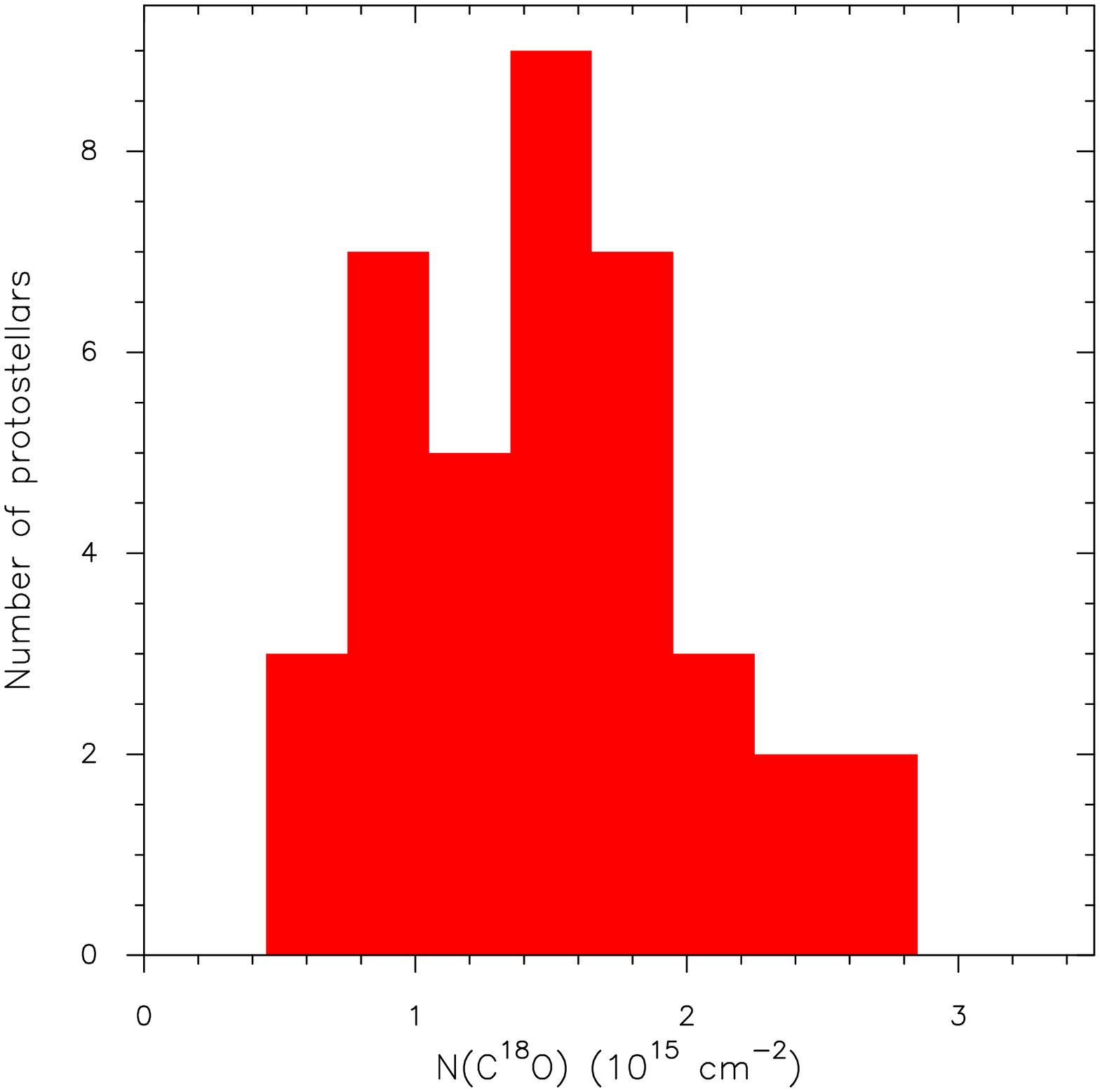}
                        \caption{}
                \end{subfigure}
                \caption{Histograms of the distributions of (a) the number of pixels with certain column density of $\rm C^{18}O(1-0)$ , (b) the number of prestellar cores with this density, and (c) the protostellar cores with this column density. The prestellar and protostellar sources clearly sample the larger \c18o \ \ column densities.
                }
                \label{fig:Histogram}
        \end{figure*}

        \subsection{Column densities and abundance ratios} \label{column-abun}
        The excitation temperature, $T_{\rm ex}$, was estimated from the peak brightness temperature of $\rm  ^{12}CO(1-0)$, by the following equation with the assumption that the $\rm  ^{12}CO(1-0)$ line is optically thick and taking into account a beam-filling factor of one \citep[e.g.][]{2010ApJ...721..686P, 2015ApJ...805...58K, 2016ApJ...826..193L}:
        \begin{equation}
        \begin{split}
        T_{\rm ex}&=\frac{hv_{\rm ^{12}CO}}{k}\bigg[\ln\bigg(1+\frac{hv_{\rm ^{12}CO}/k}{T_{\rm mb,^{12}CO}+J_v(T_{\rm bg})}  \bigg)\bigg]^{-1} \rm K \\
        &=5.53\bigg[\ln\bigg(1+\frac{5.53}{T_{\rm mb,^{12}CO}+0.818}  \bigg)\bigg]^{-1} \rm K,
        \end{split}
        \end{equation}
        where $T_{\rm mb,^{12}CO}$ is the peak intensity of $\rm  ^{12}CO(1-0)$ in units of K, $J_v(T)=\frac{hv/k}{\exp(hv/(kT))-1}$ is the effective radiation temperature \citep{1976ApJS...30..247U}, and $T_{\rm bg}=2.7$\,K is the temperature of cosmic microwave background radiation.
        The estimated excitation temperatures in the whole observed region range from 3.6 to 23.6\,K, but these values are underestimates in locations where the \co \ emission is affected by self-absorption.
        We assume that the excitation temperatures ($T_{\rm ex}$) of \s \ and \c18o \  lines have the same value as that of the optically thick \co \ line.
        The optical depth and the column densities of  $\rm ^{13}CO(1-0)$ and $\rm C^{18}O(1-0)$ were estimated using the following equations on the assumption that the cloud is in local thermodynamic equilibrium \citep[LTE; e.g.][]{1994ApJ...429..694L,1998ApJS..117..387K, 2016ApJ...826..193L}:
        
        \begin{equation}
        \tau ({\rm C^{18}O})=-\ln\bigg[1-\frac{T_{\rm mb,C^{18}O}}{5.27[J_1(T_{\rm ex})-0.166]}\bigg],
        \end{equation} 
        \begin{equation}
        \tau ({\rm ^{13}CO})=-\ln\bigg[1-\frac{T_{\rm mb,^{13}CO}}{ 5.29[J_2(T_{\rm ex})-0.164]}\bigg]
        ,\end{equation} 
        and
        \begin{equation}
        N({\rm C^{18}O}) =2.42\times 10^{14}\frac{\tau ({\rm C^{18}O}) \Delta V({\rm C^{18}O})T_{\rm ex}}{1-\exp (-5.27/T_{\rm ex})} 
        ,\end{equation} 
        \begin{equation}
        N({\rm ^{13}CO}) =2.42\times 10^{14}\frac{\tau ({\rm ^{13}CO}) \Delta V({\rm ^{13}CO})T_{\rm ex}}{1-\exp (-5.29/T_{\rm ex})}
        ,\end{equation} 
        where $J_1(T_{\rm ex})=1/[\exp(5.27/T_{\rm ex})-1]$,  $J_2(T_{\rm ex})=1/[\exp(5.29/T_{\rm ex})-1]$, and $\Delta V$ is the FWHM in \kms.
        The abundance ratio, $X{\rm (^{13}CO)}/X{\rm (C^{18}O)}$, is equivalent to the $N{\rm (^{13}CO)}/N{\rm (C^{18}O)}$ column density ratio.

        In the regions where the \co \ emission is affected by self-absorption, the column densities of \s \ and \c18o \ \ are effected by underestimates of $T_{\rm ex}$ from the \co \ data.
        This leads to underestimation of the column densities of \s \ and \c18o \ \ in those areas.
        The resulting ranges of $N({\rm ^{13}CO})$ and $N({\rm C^{18}O})$ are 3.8$\times10^{14}$--5.5$\times10^{16}$\,cm$^{-2}$ and 2$\times10^{14}$--3.4$\times10^{15}$, respectively, and the resulting range of the abundance ratio of R$_{13/18}$ is 1.02--41.9. 
        A map of the column density of $\rm C^{18}O(1-0)$ is represented in Figure \ref{fig:N}. 
        
        The column density of $\rm C^{18}O(1-0)$ versus the number of pixels with $\rm C^{18}O(1-0)$ is displayed in the diagrams of Figure \ref{fig:Histogram} together with the observed column densities at the location of the prestellar and protostellar cores. 
        The cores sample locations with increasing \c18o \ \ column density confirming that star formation first happens in regions with the highest column densities.
        Most of the $\rm C^{18}O(1-0)$ detection points correspond to the column densities in the range 0.5$\times10^{15}$ $\thicksim1.6$$\times10^{15}$\,cm$^{-2}$ (see panel (a)).  
        The prestellar and protostellar cores correlate well with the \c18o \ \ column density.
        The  the  majority of the prestellar cores  column densities in the range of 0.5$\times10^{15}$$\thicksim$2$\times10^{15}$\,cm$^{-2}$ while the locations of the  protostellar cores coincide with the column densities in the range of 0.5$\times10^{15}$$\thicksim$2.8$\times10^{15}$\,cm$^{-2}$.
        \begin{figure*}[hbt!]
                \centering
                \begin{subfigure}[b]{.49\linewidth}
                        \includegraphics[width=\linewidth]{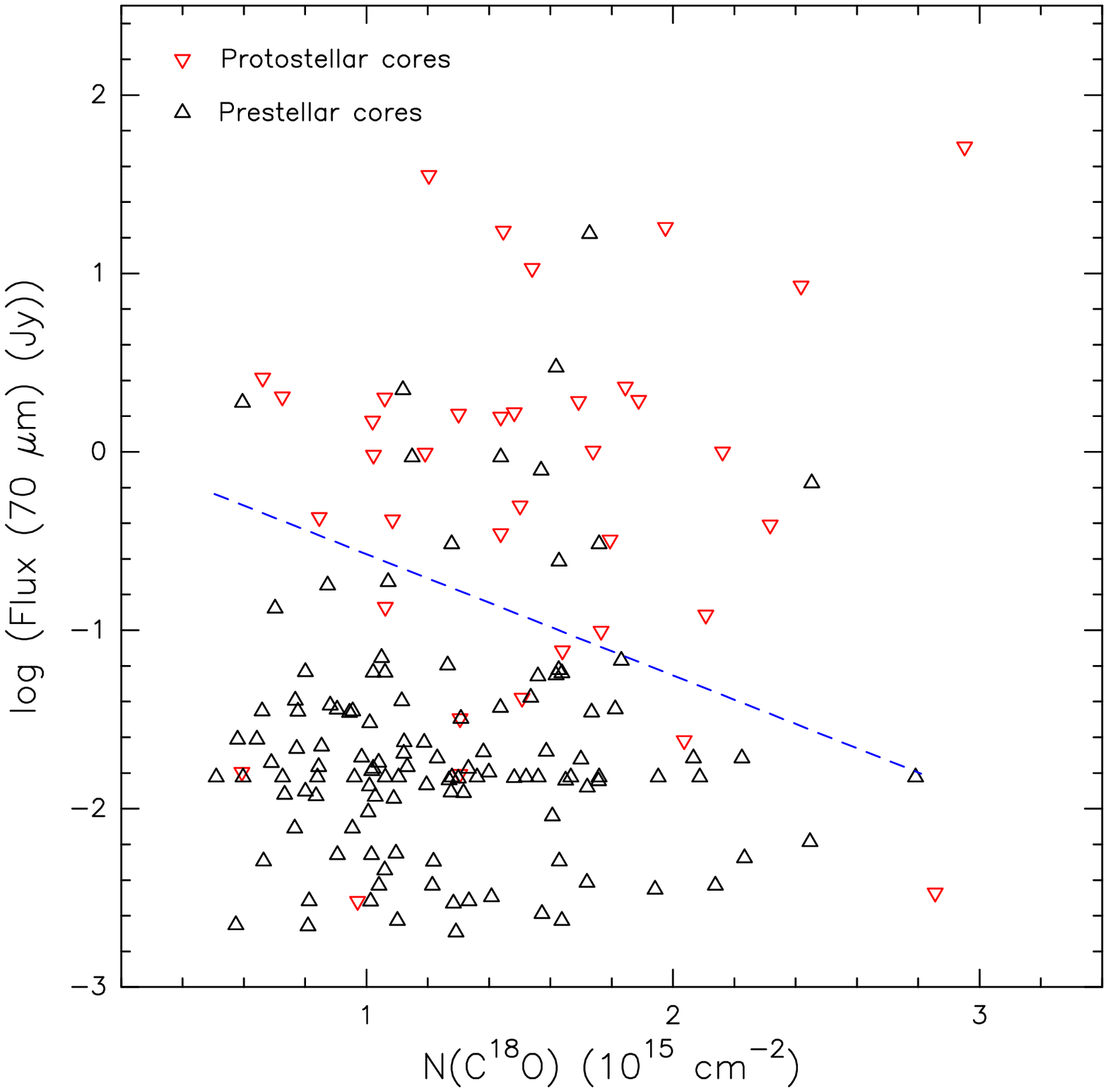}
                        \caption{}
                \end{subfigure}
                \begin{subfigure}[b]{.49\linewidth}
                        \includegraphics[width=\linewidth]{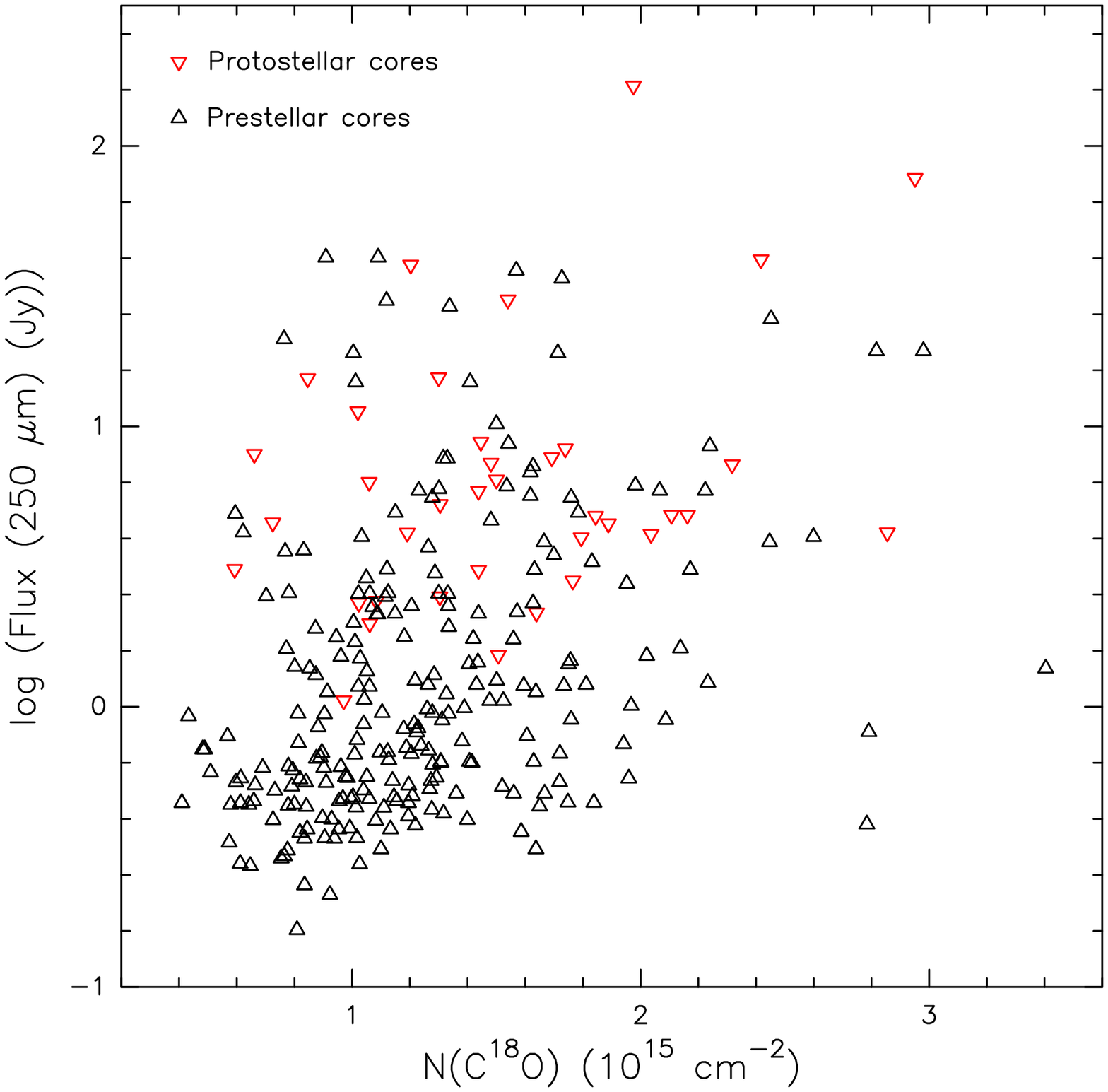}
                        \caption{}
                \end{subfigure}
                \caption{Column density of \c18o \  vs. the peak flux densities of protostellar (red triangles) and prestellar (black triangles) cores at (a) 70 $\mu$m \ and (b) 250 $\mu$m \ selected from \cite{2015A&A...584A..91K}. The number of prestellar cores in panel (a) is less than the total number because some have no 70 $\mu$m \ flux. A clear separation is found between prestellar and protostellar cores in panel (a) (shown with a dotted line). Panel (b) shows an increasing trend of the 250 $\mu$m flux density with the \c18o \  column density suggesting that star formation is more advanced at higher column densities and that an increasing volume is affected by the ongoing star formation.
                }
                \label{fig:flux-N}
        \end{figure*}
        \begin{figure}[hbt!]
                \centering 
                \includegraphics[width=\linewidth]{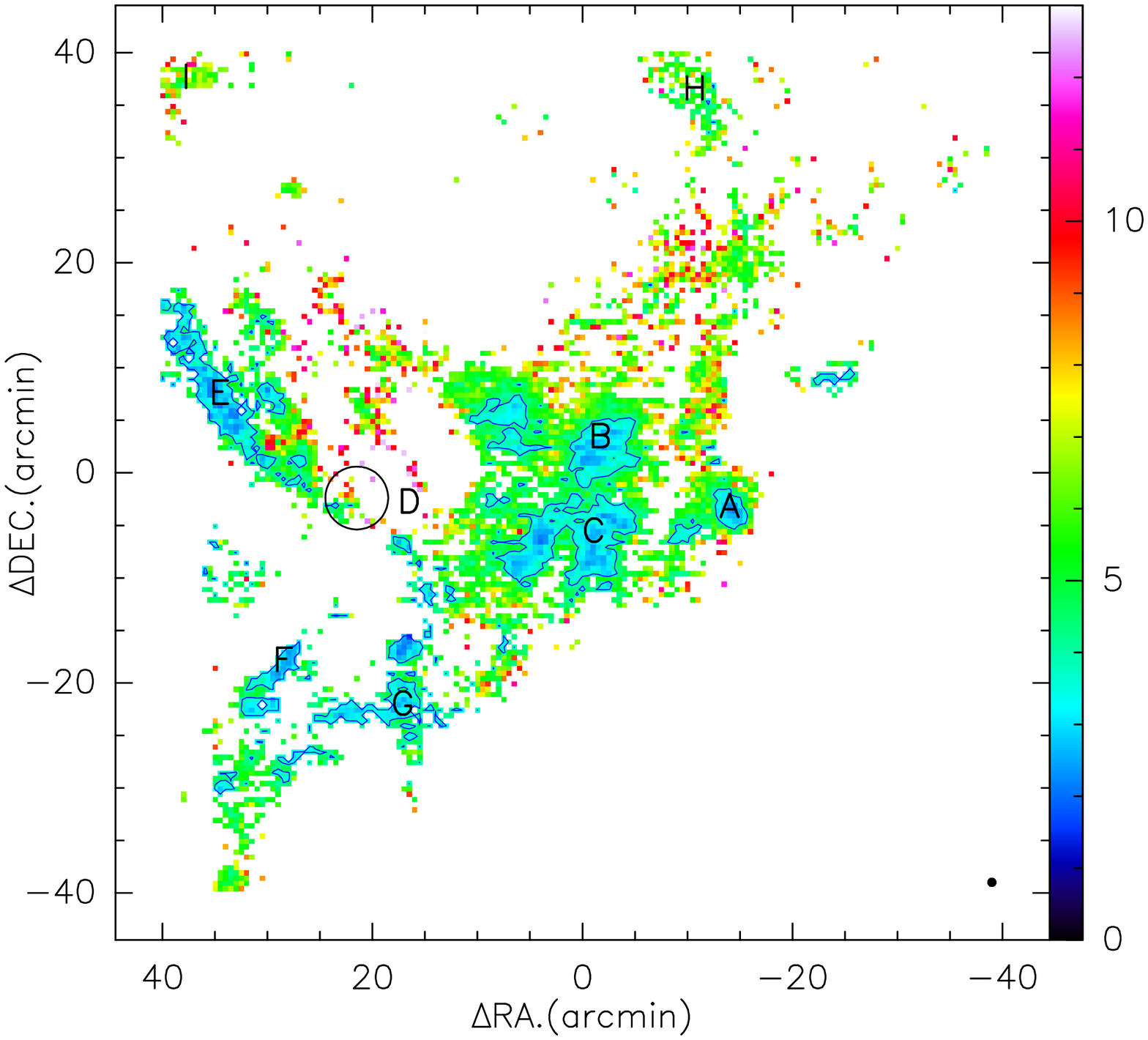}
                \caption{Map of the abundance ratio $ X_{\rm ^{13}CO}$/$X_{\rm C^{18}O}$. The contours show the value of $ X_{\rm ^{13}CO}$/$X_{\rm C^{18}O}$=4. The letters A to I present the selected regions where the abundance ratio is lower (A, B, C, E, F, and G) and much higher (region D), and the outskirts (H and I).
                        The black circle presents the location of the W40 H\,\Rn2 \ region (radius $\thicksim$ 3\arcmin) and the black filled circle in the bottom-right illustrates the beam size of 60\arcsec. The clipping level for the \s \ and \c18o \ data is the signal-to-noise ratio greater than five. High abundance-ratio values are distributed only in region D and reach up to 42.}
                \label{fig:ratio}
        \end{figure}
        
        \subsection{Far-infrared distribution}
        
        The peak flux density at the prestellar and protostellar cores at 70 $\mu$m \ and 250 $\mu$m is presented versus the column density of $\rm C^{18}O(1-0)$ diagrams in Figure \ref{fig:flux-N}.
        The 70 $\mu$m  diagram (panel (a)) shows a clear separation indicated with a dotted line between the prestellar and protostellar core distributions; most of prestellar cores have a lower flux density while most of the protostellar cores have a higher flux that may increase slightly at higher column densities. 
        However, excluding the uncertainty on the $\rm N(C^{18}O)$ emission, the 250 $\mu$m distribution in panel (b) shows an increasing trend with column density suggesting that star formation is more advanced at higher column densities and that an increasing volume is affected by the ongoing star formation.
        Star formation processes are slower in regions of lower column density and more time is required to heat the surrounding environment.
        %At the higher column density, the stellar evolution is faster because the accretion rate unto prestellar core is abundant; therefore, this will become the more advanced stage and will heat their surrounding materials earlier.  
        In addition, the number of protostellar (or prestellar) cores may be larger because of fragmentation which will accelerate the heating process. 
        %Protostellar cores are the first stage for stars to be formed, and they will have succeeded to heat their surrounding environment. 
        %Hence is why most of them have high magnitudes of flux.
        The number of cores with very high column density is much smaller than for low-column-density cores, which agrees with the histogram of the stellar core number versus the \c18o \  column density in Figure \ref{fig:Histogram}.

        \subsection{Selective photodissociation} 
        \label{SD}
        A map of the abundance ratio $ X_{\rm ^{13}CO}$/$X_{\rm C^{18}O}$ is presented in Figure \ref{fig:ratio}. The contours show the value of $ X_{\rm ^{13}CO}$/$X_{\rm C^{18}O}$=4.
        We selected nine unique regions (A-I) in the map where the abundance ratio has a  lower value (A, B, C, E, F, and G) , one region with a much higher value (region D), and two regions on the outskirts with medium values (H and I).
        The abundance ratio located in region D next to the W40 H\,\Rn2 \ region is as high as 42, which is significantly higher than the value of 5.5 seen in the Solar System \citep{1989GeCoA..53..197A}. 
        Region D is located next to the W40 (northwest side), region B is Serpens South, and region G was identified as a distinct star formation region by \cite{2019ApJ...874..172K}.
        The far-ultraviolet (FUV) emission with sufficient energy to photo-dissociate \co \  promptly becomes optically thick as it penetrates molecular clouds.
        On the contrary, \c18o \ self-shielding is relatively less significant because of its low abundance and the shift of its absorption lines. 
        Hence, this infers that the FUV emission with higher energy than the \c18o \ \ dissociation level will penetrate deeper into a molecular cloud and will also increase the abundance ratio of \s \  and \c18o \ , \xab , at the instance when \c18o \ self-shielding is not yet dominant. 
        When both self-shielding effects of \s \  and \c18o \ \ become significant, the abundance ratio will decrease towards its intrinsic value, which is derived from  $\rm ^{12}C$,$\rm ^{13}C$, $\rm ^{16}O$, and $\rm ^{18}C$ abundances \citep{1988ApJ...334..771V, 2007A&A...476..291L, 2014A&A...564A..68S}.       
        We can therefore confirm that selective FUV photodissociation of \c18o \ indeed occurs.

        \begin{figure*}[hbt!]
                \centering 
                \includegraphics[width=\linewidth]{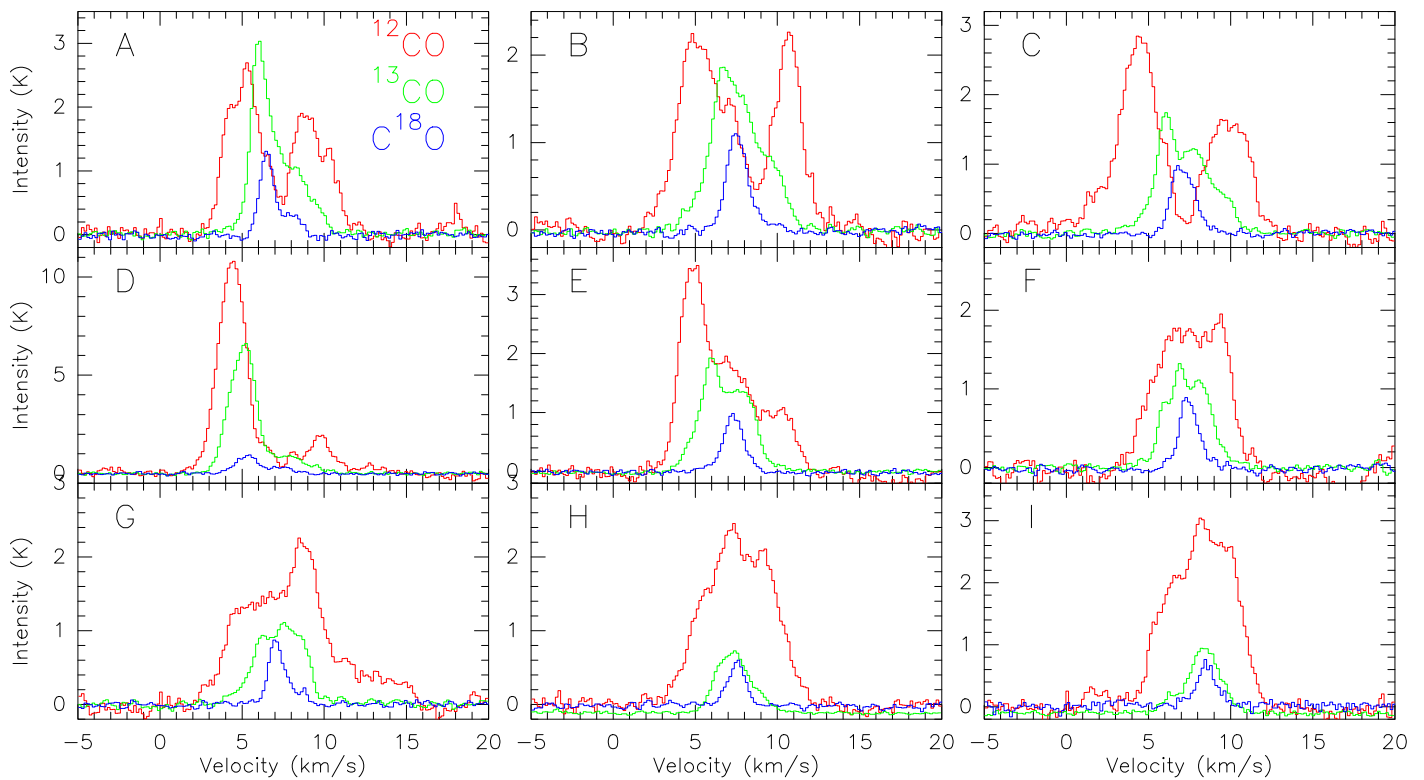}
                \caption{Averaged spectra of $\rm ^{12}CO(1-0)$ (red), $\rm ^{13}CO(1-0)$ (green), and $\rm C^{18}O(1-0)$ (blue) emission at the selected regions from Figure \ref{fig:ratio}. The name of the selected region is presented in the top-left corner of each panel.}
                \label{fig:spec}
        \end{figure*}

        Figure \ref{fig:spec} presents the averaged spectra of $\rm ^{12}CO(1-0)$, $\rm ^{13}CO(1-0),$ and $\rm C^{18}O(1-0)$ emission in the nine selected regions. 
        The central component of the broader \co \ can clearly be seen to be depleted by self-absorption. 
        However, the spectrum shapes of \s \ and \c18o \ appear different from that of \co .

        \begin{figure*}[hbt!]
                \centering 
                \includegraphics[width=\linewidth]{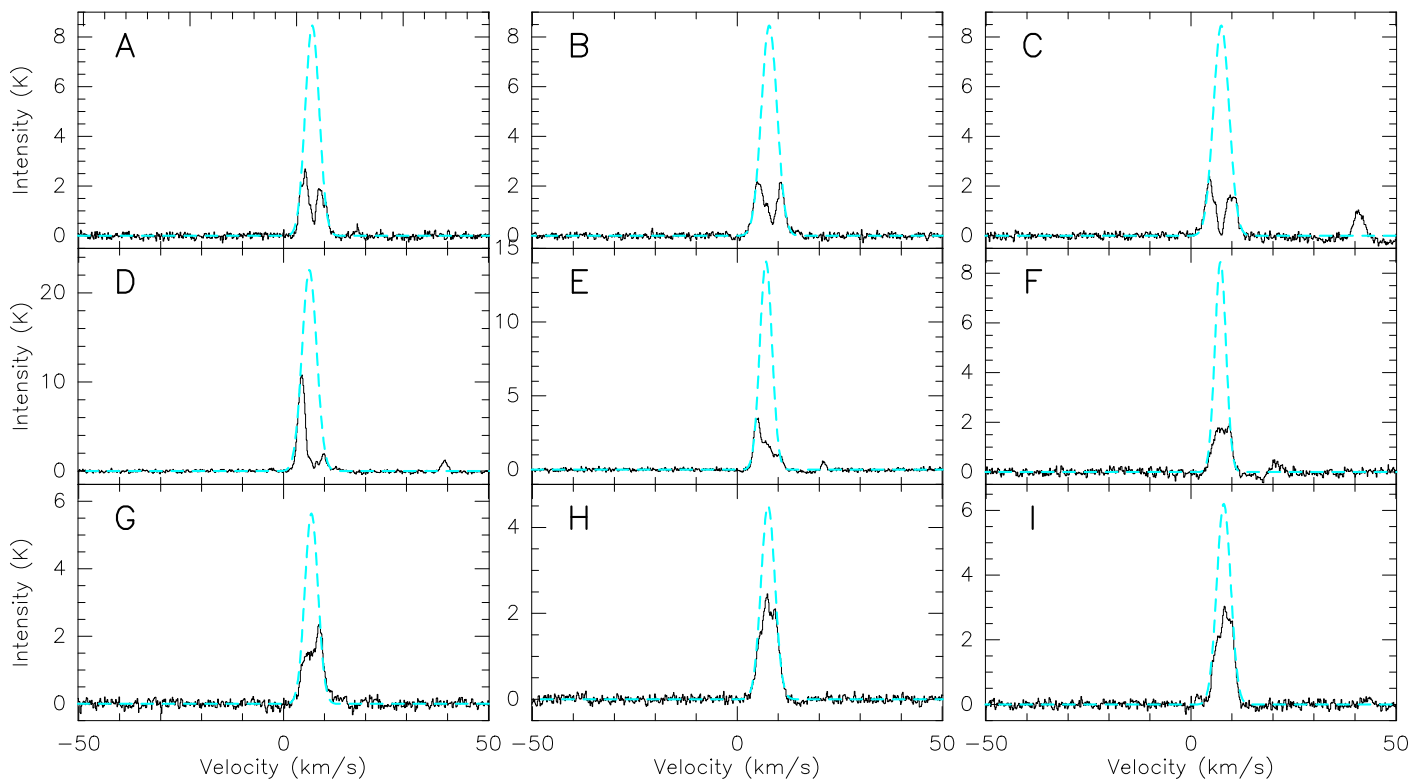}
                \caption{Examples of  the \co \ self-absorption correction. The blue dashed line presents the best single Gaussian fitting for the original \co \ profile using the remaining wings of the emission line. The name of the selected region is presented in the top-left corner of each panel.}
                \label{fig:correction}
        \end{figure*}
        
        \begin{figure}[hbt!]
                \centering 
                \includegraphics[width=\linewidth]{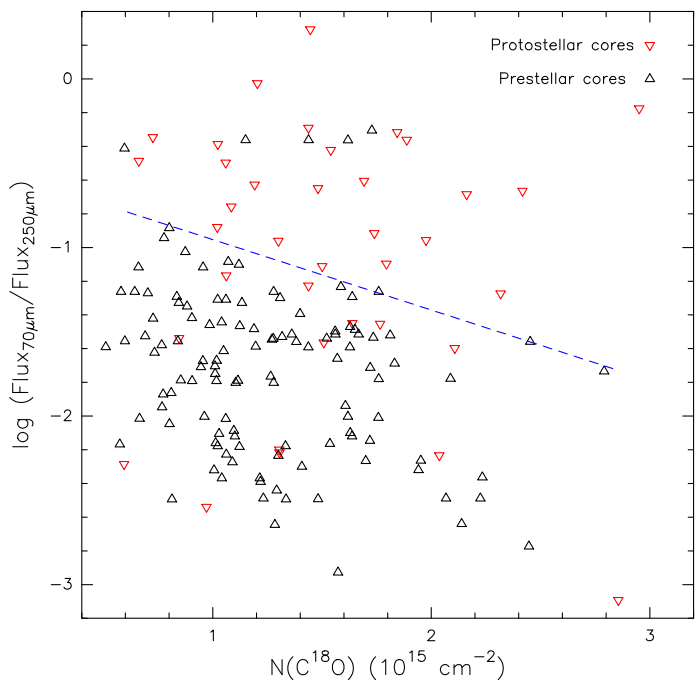}
                \caption{Column density of \c18o \  vs. the peak flux density ratio of the bandwidth of 70 $\mu$m \ and 250 $\mu$m \ of the                   protostellar (red triangles) and prestellar (black triangles) cores selected from \cite{2015A&A...584A..91K}. The dotted line indicates               the dividing line between prestellar and protostellar cores and shows the gradual change of the ratio.}
                \label{fig:ratio6}
        \end{figure}
        
        \begin{figure}[hbt!]
                \centering 
                \includegraphics[width=\linewidth]{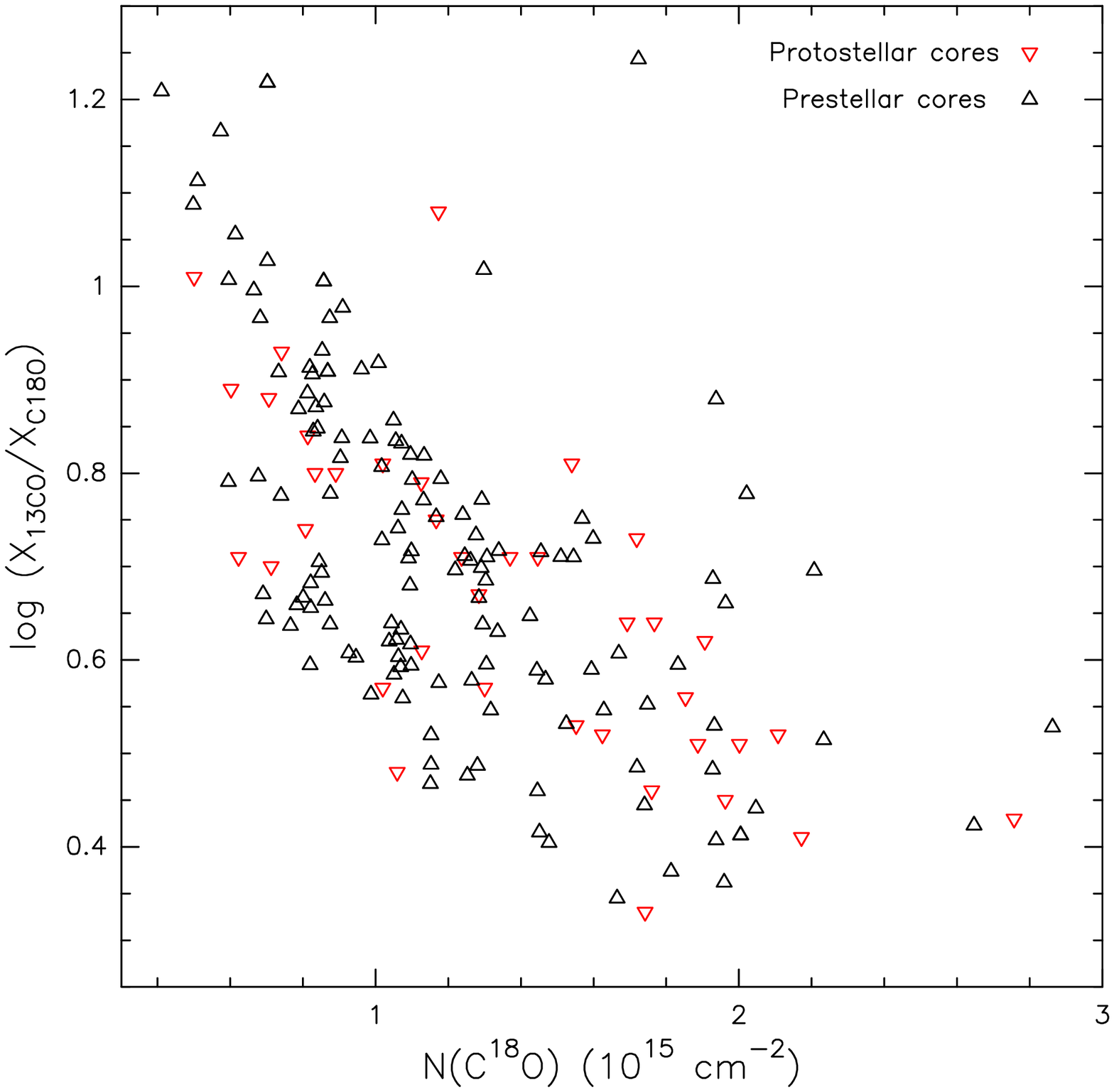}
                \caption{Column density of \c18o \  vs. the abundance ratio $ X_{\rm ^{13}CO}$/$X_{\rm C^{18}O}$ in the protostellar (red triangles) and prestellar (black triangles) cores.}
                \label{fig:N-X}
        \end{figure}

        \subsection{Integrated properties of the distinct regions}
        
        The molecular hydrogen gas masses ($M_{H_2}$) of the selected regions  A to I were derived using equations from \cite{2013MNRAS.436..921T}.
        In order to estimate the original mass of the selected regions, we fitted the profile of the \co \  emission line with a Gaussian using the line wings of the self-absorbed \co \ lines (see Figure \ref{fig:correction}).
        
        We compute the \co \ luminosity as follows:
        \begin{equation}
        L_{CO}=I_{avg} \times N_{pix} \times 23.5 \times (D\times \Delta_{pix})^2,
        \end{equation} 
        where $L_{CO}$ is the \co \ luminosity in $\rm K \ kms^{-1} pc^2$ , $I_{avg}$ is the average \co \ intensity ($\rm K \ kms^{-1}$ ) obtained from the integrated intensity
        map on a main-beam temperature scale, $N_{pix}$ is the number of pixels included, D is the distance to the sources in megaparsec, and  $\Delta_{pix}$ is the pixel size in arcseconds. The mass of the molecular gas is computed using:
        \begin{equation}
        M_{H_2}=1.6\times 10^{-20}\times L_{CO} \times X_{CO} \ \ \ (M_\odot\ ),
        \end{equation} 
        where $\rm X_{CO}=2 \times 10^{20} (K \ kms^{-1})^{-1}$ \citep{1988A&A...207....1S}.
        
        The average fluxes of the infrared 70 $\mu$m \ and 250 $\mu$m \ images of the selected  regions (see Table \ref{tab:1}) were obtained from the Herschel Gould Belt Survey Archive.
        The average 70 $\mu$m \ flux is higher in regions A, B, C, D, E, F, and G, with the highest flux being $\thicksim$ 2172 MJy/sr in region D next to the W40 H\Rn2 \ region. However, the average 70 $\mu$m \ flux is significantly lower in regions H and I. This means that regions A, B, C, D, E, F, and G are enhanced, and that the star-forming processes therein are more active than those in regions H and I. 
       The integrated properties of the distinct regions are summarised in Table \ref{tab:1}. The columns have the following meaning:
        Column 1: The distinct region,
        Column 2 and 3: The right ascension and declination (J2000) of the centre position,
        Column 4: The radius of the region,
        Column 5: The average flux of Herschel 70 $\mu$m \ image,
        Column 6: The average flux of Herschel 250 $\mu$m \ image,
        Column 7: The original mass based on the best Gaussian fitting, and
        Column 8: The original luminosity of the \co \ based on the best Gaussian fitting.

        \begin{table*}[hbt!]
                \caption{Integrated properties of the distinct regions}
                % title of Table
                \label{tab:1}
                % is used to refer this table in the text
                \centering
                % used for centering table
                \begin{tabular}{c|ccccccc}
                        % centered columns (4 columns)
                        \hline\hline
                        % inserts double horizontal lines
                        Reg.&RA & DEC & Area & $\rm F_{70 \mu m}$& $\rm F_{250 \mu m}$ &$\rm M_{H_2}$ &$\rm L_{^{12}CO}$\\
                        &J2000&J2000& $\rm pc^2$ & MJy/sr & MJy/sr & \(M_\odot\) & $\rm K \ kms^{-1} pc^{2}$ \\
                        (1)&(2)&(3)&(4)&(5)&(6)&(7)&(8)\\
                        % table heading
                        \hline
                        A    &    18:29:07.40   &   -02:05:28    &    0.97  &  80.97    &  416.56     & 33.48   &    10.46      \\
                        B    &    18:29:57.80   &   -02:00:04    &    1.16  &  187.07   &  599.29     & 47.94   &    14.98      \\
                        C    &    18:29:58.78   &   -02:10:37    &    1.09  &  77.51    &  375.55     & 41.09   &    12.84      \\
                        D    &    18:31:03.24   &   -02:05:28    &    1.09  &  2171.54  &  1,319.11   & 97.17   &    30.36     \\
                        E    &    18:31:57.15   &   -02:00:14    &    1.16  &  236.48   &  830.95     & 62.15   &    19.42      \\
                        F    &    18:32:20.73   &   -02:26:07    &    0.91  &  76.29    &  249.23     & 26.79   &    8.37       \\
                        G    &    18:31:09.80   &   -02:25:22    &    0.97  &  77.14    &  268.46     & 25.98   &    8.12       \\
                        H    &    18:29:21.57   &   -01:25:00    &    0.97  &  8.27     &  75.84      & 14.61   &    4.57       \\
                        I    &    18:32:33.50   &   -01:24:41    &    0.59  &  1.00     &  62.50      & 6.04    &    1.89       \\
                        \hline
                \end{tabular}
            \begin{tablenotes}
            	\small
            	\item Note: Column 1: The distinct region,
            	Column 2 and 3: The right ascension and declination (J2000) of the centre position,
            	Column 4: The radius of the region,
            	Column 5: The average flux of Herschel 70 $\mu$m \ image,
            	Column 6: The average flux of Herschel 250 $\mu$m \ image,
            	Column 7: The original mass based on the best Gaussian fitting, and
            	Column 8: The original luminosity of the \co \ based on the best Gaussian fitting.
            \end{tablenotes}
        \end{table*}

        \begin{figure*}[hbt!]
                \centering
                \begin{subfigure}[b]{.485\linewidth}
                        \includegraphics[width=\linewidth]{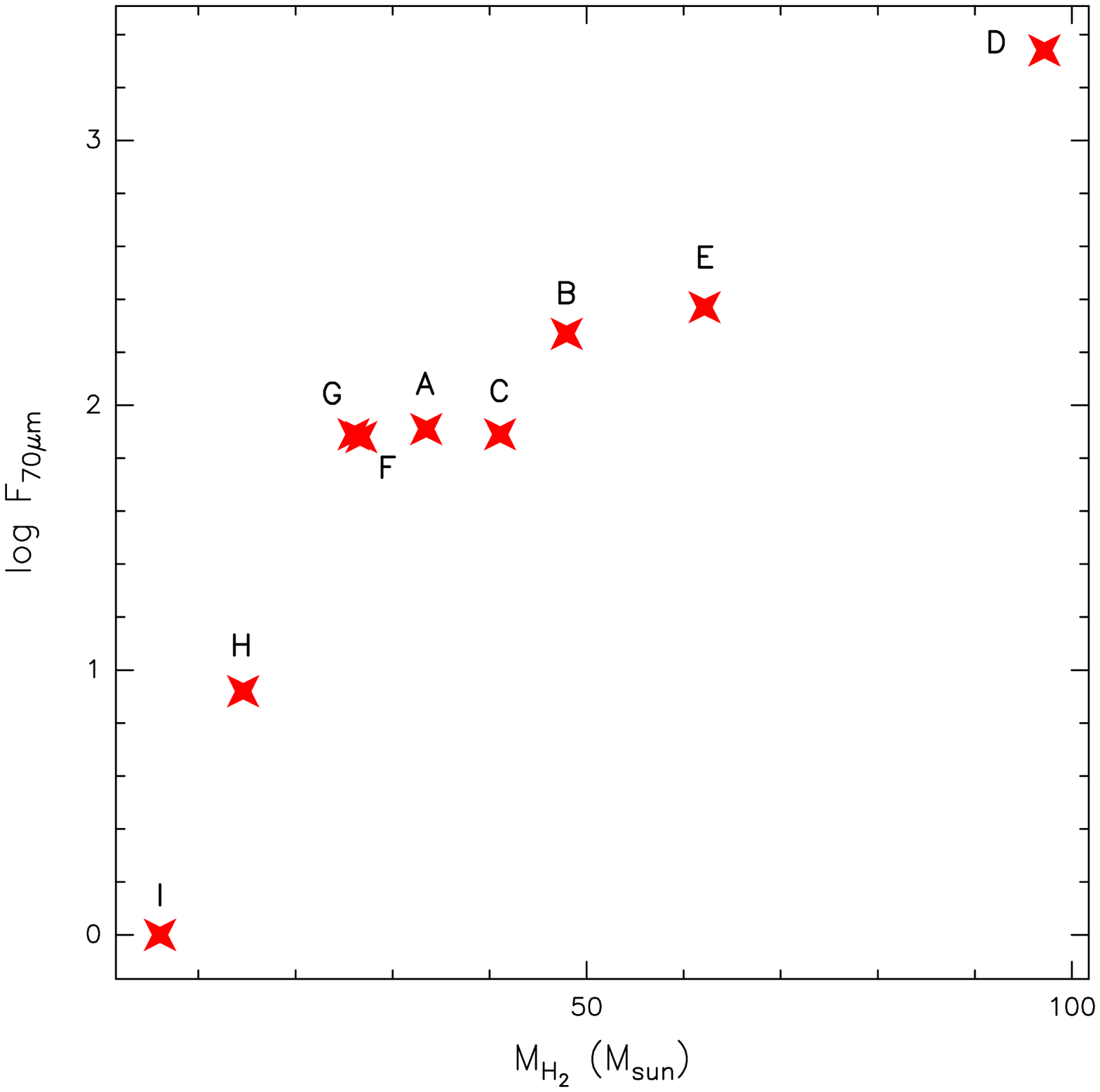}
                        \caption{}
                \end{subfigure}
                \begin{subfigure}[b]{.50\linewidth}
                        \includegraphics[width=\linewidth]{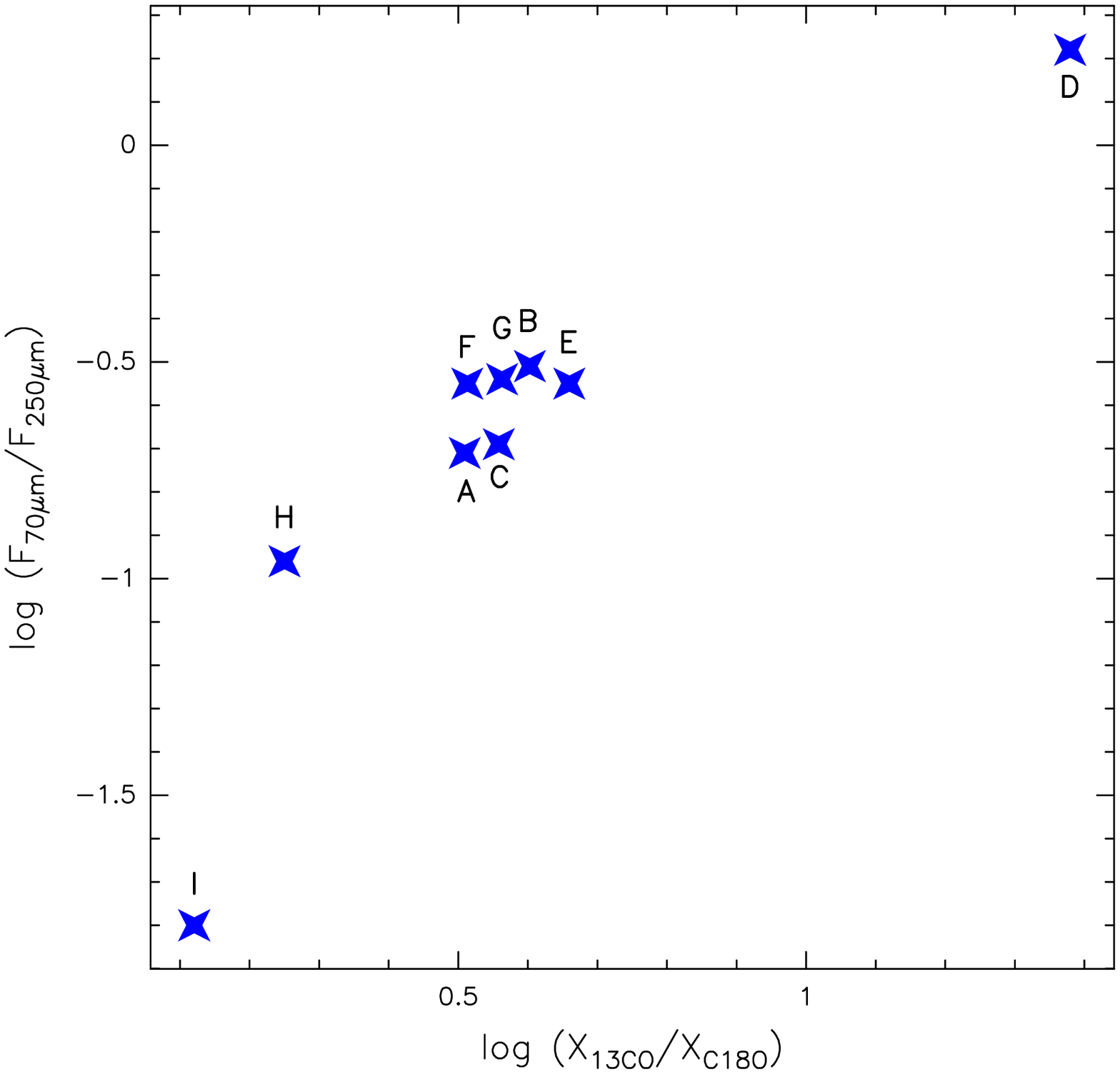}
                        \caption{}
                \end{subfigure}
                \caption{(a) Mass-corrected self-absorption vs.  70 $\mu$m \ flux, and (b) the ration $\rm X_{^{13}CO}/X_{C^{18}O}$, equivalent to the dissociation rate of the \c18o , vs. the flux ratio $\rm F_{70           \mu m}/F_{250 \mu m}$ of the distinct regions. Selected regions are marked with the letters A to I. The average fluxes of the infrared 70             $\mu$m \ and 250 $\mu$m\ images of the distinct regions in both panels were obtained from the data of the Herschel Gould Belt                Survey Archive.
                }
                \label{fig:perc}
        \end{figure*}

        \begin{figure}[hbt!]
                \centering 
                \includegraphics[width=\linewidth]{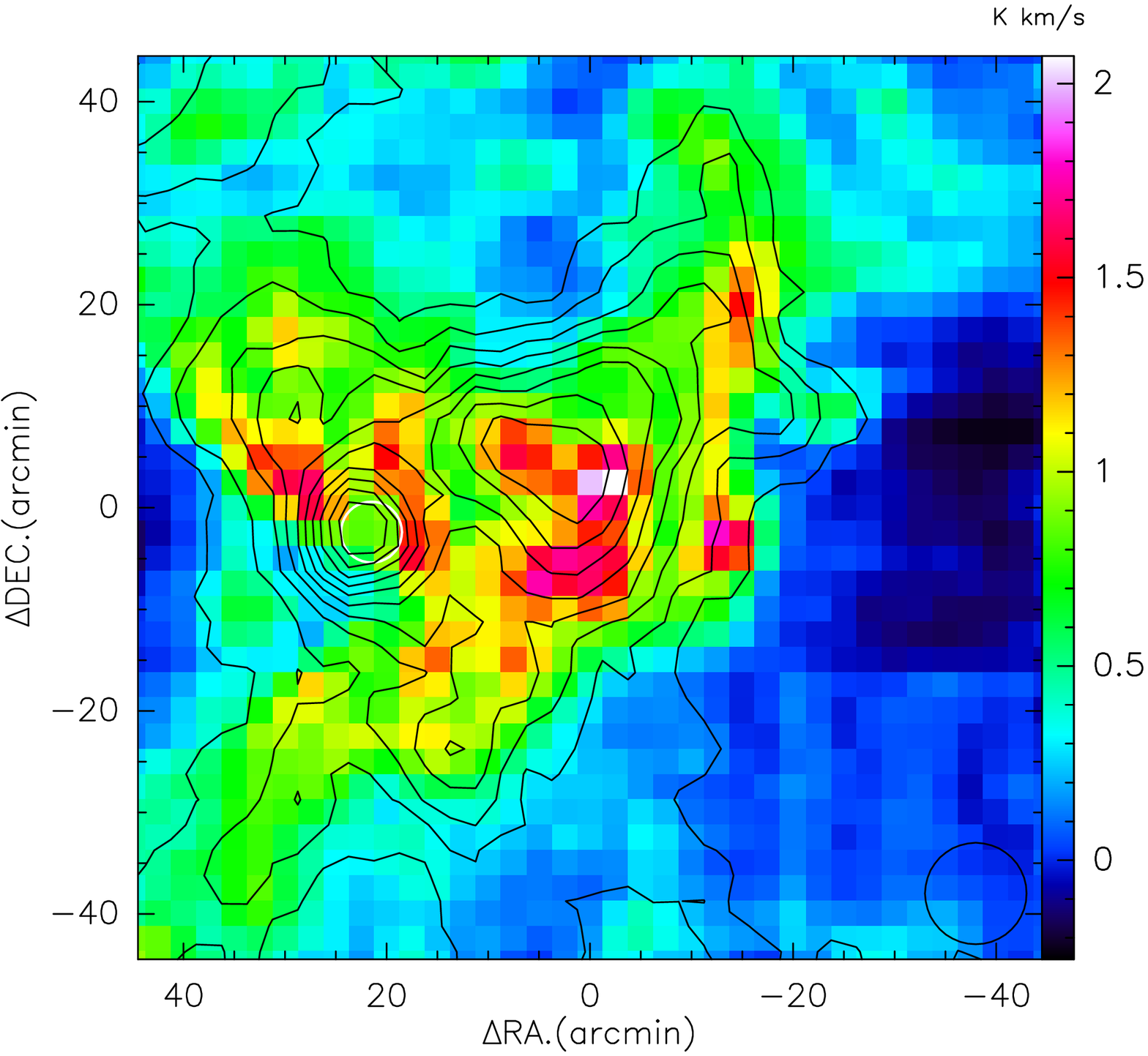}
                \caption{Map of the $\rm C^{18}O(1-0)$ integrated intensity superposed on the contours of $\rm H_2CO$ absorption toward the                         AMC.  The $\rm C^{18}O(1-0)$ and $\rm H_2CO$ data were smoothed to the same effective resolution and cell size of 10\arcmin \ and 2.5 \arcmin, respectively. Contour levels of the $\rm H_2CO$ intensity map                   are $-$0.4 to $-$1.8 in steps of $-$0.15\,K \kms. The white circle presents the location of the W40 H\,\Rn2 \ region (radius $\thicksim$ 3\arcmin) and the black circle in the bottom-right illustrates the resolution of 10\arcmin.}
                \label{fig:h2co-c18o}
        \end{figure}

        \subsection{Correlating the infrared and photo-dissociation data}
        
        The heating of the interstellar dust by the star formation processes first results in 250 $\mu$m \ infrared emission and at a later stage produces 70 $\mu$m \ emission  \cite[eg.][]{2015MNRAS.448..135B}.  
        Therefore, the 70/250 $\mu$m \ ratio could be an indicator of the evolutionary age of the star formation process and the state of the interstellar dust. 
        On the other hand, photo-dissociation is clearly a time-dependent effect resulting from the UV exposure of the molecular gas and the photo-dissociation rate also reflects the evolutionary age of the selected distinct regions.

        The peak flux density at the prestellar and protostellar cores at a wavelength of 70 $\mu$m \ versus the column density of $\rm C^{18}O(1-0)$ \  reveals a separation between the prestellar and protostellar core distributions (Figure \ref{fig:flux-N}a). 
        The variation of the 250 $\mu$m \ emission data suggests a flux density increase at higher  \c18o \  column densities, such that the more intense star formation is able to heat a larger volume of the surrounding dust, which is also time dependent (Figure \ref{fig:flux-N}b).
        %The difference between 70 $\mu$m \ and 250 $\mu$m \ flux distributions confirms that the stellar cores first appear in 250 $\mu$m \ emission and maybe later in 70 $\mu$m \ emission. 
        Figure \ref{fig:ratio6} displays the column density of \c18o \  versus the peak flux density ratio $\rm F_{70 \mu m}/F_{250 \mu m}$ of protostellar and prestellar cores. 
        Excluding some improperly distributed stellar cores, most evolved protostellar cores are in the upper part, and the youthful prestellar cores are in the lower part in the diagram.
        However, the dividing line between the flux ratio decreases with an increasing \c18o \  column density because at higher \c18o \ columns the star formation has time and energy to heat larger volumes of dust.
        The flux ratio $\rm F_{70 \mu m}/F_{250 \mu m}$ diagram confirms that the \c18o \  column density is related to the evolutionary age of the cores.

        The abundance ratio of $ X_{\rm ^{13}CO}$/$X_{\rm C^{18}O}$ for the prestellar and protostellar locations plotted against the \c18o \  column density shows a similar distribution for prestellar and protostellar cores, which indicates that the effects of photodissociation only depend on the presence of star formation at this \c18o \  column density (Figure \ref{fig:N-X}). 
        The $ X_{\rm ^{13}CO}$/$X_{\rm C^{18}O}$ abundance ratio decreases with increasing  \c18o \  column density indicating an enhancement of \c18o \ in the interstellar medium. The strong variation of the abundance ratio with the \c18o \  column density shows that the more massive and fragmented star formation has a strong effect on the surrounding area.

        The actual mass of the star formation region containing young stellar objects plays a critical role in distinguishing between evolutionary paths and their timescales. 
        In order to depict the evolutionary path of star formation regions in the AMC, the original mass estimated from the 
        reconstructed \co \ emission line in the nine selected regions is plotted against the 70 $\mu$m \ flux density in Figure \ref{fig:perc}a.
        This diagram shows that the regions lie along a string of points indicating five stages of evolution.
        Region D, next to the W40 H\Rn2 , belongs to the oldest evolutionary stage with the highest amount of photo-dissociated mass ($\thicksim$ 74 \msun \ ) , and with the highest 70 $\mu$m \ flux density. 
        The regions B and E form the next oldest stage of evolution, where region B is the Serpens South region and region E is a filament located southeast of the W40 H\Rn2 . Region E has lost more mass to photo-dissociation than region B.
        Regions G, F, A, and C resemble regions of an earlier stage of evolution, where region C has lost a little more mass than the others because it is larger. 
        The youngest star formation region H is just starting to show up in 70 $\mu$m \ images. Region I is the youngest region located on the outskirts of the cloud; it does not yet show any 70 $\mu$m \ flux and has the least estimated mass of all the selected regions.  
        
        As UV radiation selectively dissociates less-abundant CO isotopologues more effectively than more-abundant ones \citep{1988ApJ...334..771V, 2007A&A...476..291L, 2014A&A...564A..68S} the ratio 
        $\rm X_{^{13}CO}/X_{C^{18}O}$ is equivalent to the dissociation rate of the \c18o . 
        A similar evolutionary sequence is found when plotting the ratio $\rm X_{^{13}CO}/X_{C^{18}O}$ against the flux ratio $\rm F_{70 \mu m}/F_{250 \mu m}$ of the selected regions in Figure \ref{fig:perc}b. 
        However, the mid-range evolutionary regions cluster together in this case, possibly because of  inaccuracies in estimating the abundance ratio. 
        The two diagrams of Figure \ref{fig:perc} present the original mass, and the rate of photodissociation combined with the infrared properties are useful to determine the evolutionary age of the star-forming regions.

        % \subsection{
        % Influence of the uncertainties in the beam filling factor and the excitation temperature on our derived abundance ratio $ X_{\rm ^{13}CO}$/$X_{\rm C^{18}O}$}
        % In Sect. \ref{column-abun}, we derived the optical depths and column densities of \s \ and \c18o \ assuming the beam filling factors, $\phi_{^{13}CO}$ and $\phi_{C^{18}O}$ are 1.0, and the excitation temperatures ($T_{\rm ex}$) of \s \ and \c18o \  lines have the same value as that of the optically thick \co \ line. In this subsection, we evaluate the influence of the beam filling factor and the excitation temperature on the derived physical properties.
        
        \subsection{Uncertainties on the abundance ratio $ X_{\rm ^{13}CO}/X_{\rm C^{18}O}$ introduced by 
                the excitation temperature and the beam-filling factor }
        
        \c18o \ and  \s \ beam-filling factors were assumed to be unity when estimating  their column densities and optical depths in  Sect. \ref{column-abun}. Also, the excitation temperatures ($T_{\rm ex}$) for all three lines \c18o , \s, \ and \co \ were presumed to have the same values, as derived from the optically thick line \co. 
        
        \begin{table*}[hbt!]
                \caption{Column densities and abundance ratio $ X_{\rm ^{13}CO}$/$X_{\rm C^{18}O}$ of $\rm ^{13}CO$ and $\rm C^{18}O$}
                % title of Table
                \label{tab:factor}
                % is used to refer this table in the text
                \centering
                % used for centering table
                \begin{tabular}{c|cccc}
                        % centered columns (4 columns)
                        \hline\hline
                        % inserts double horizontal lines
                        Molecule    & $T_{\rm ex}$           & Beam-filling factor &  Column density    &  $ X_{\rm ^{13}CO}$/$X_{\rm C^{18}O}$    \\
                        \hline
                        $\rm ^{13}CO$   & $T_{\rm ex}\rm (^{12}CO)$  & 1   &  0.38--55.34$\times10^{15}$cm$^{-2}$ &  \multirow{2}{*}{1.02--41.9}  \\
                        $\rm C^{18}O$   & $T_{\rm ex}\rm (^{12}CO)$  & 1   &  0.21--3.42$\times10^{15}$cm$^{-2}$  &                                \\
                        \hline
                        $\rm ^{13}CO$   & $T_{\rm ex}\rm (^{12}CO)$  & 0.6 &  0.77-70.41$\times10^{15}$cm$^{-2}$  &   \multirow{2}{*}{0.99--40.1} \\
                        $\rm C^{18}O$   & $T_{\rm ex}\rm (^{12}CO)$  & 0.6 &  0.43--15.11$\times10^{15}$cm$^{-2}$ &                                          \\
                        \hline
                        $\rm ^{13}CO$   & 30\,K                   & 1   &  0.72--49.61$\times10^{15}$cm$^{-2}$ &   \multirow{2}{*}{1.39--32.88} \\
                        $\rm C^{18}O$   & 20\,K                   & 1   &  0.28--4.17$\times10^{15}$cm$^{-2}$  &                                \\
                        \hline
                        $\rm ^{13}CO$   & 30\,K                   & 0.6 &  0.14--16.4$\times10^{16}$cm$^{-2}$  &   \multirow{2}{*}{1.38--32.88}  \\
                        $\rm C^{18}O$   & 20\,K                   & 0.6 &  0.57--9.21$\times10^{15}$cm$^{-2}$  &                                 \\
                        \hline
                \end{tabular}
        \end{table*}

        % We can estimate the beam filling factor by the equation
        % \citep{1998A&A...337..275N,2006ApJS..162..161K}:
        Calculation of the beam-filling factor is obtained through equation (8): 
        \begin{equation}
        f =\frac{\theta ^2_{\rm s}}{\theta ^2_{\rm s}+\theta ^2_{\rm beam}}
        ,\end{equation}  
        where  $\theta_{\rm beam}$ is the beam size and $\theta_{\rm s}$ is source diameter.
        The \s \ and \c18o \ data have an effective beam size of 60\arcsec corresponding to 0.124\,pc at the distance of the AMC.
        \c18o \ emission is deemed as a tracer of dense cores, filaments, and clumps. In comparison, the \s \ emission will additionally trace extended components which could be seen via integrated intensity maps \citep[][]{2014A&A...564A..68S}.
        The nearest low-mass star forming region observations toward the Taurus cloud traced via \c18o \ have dense cores of 0.1 pc in diameter \citep{1996ApJ...465..815O}.
        Moreover, the constant encounter of parsec-scale filaments with narrow widths of typically 0.1 pc within molecular clouds were discovered through Herschel micrometre emission observations \citep{2011A&A...529L...6A,2013A&A...550A..38P}.
        This infers that traced \s \ and \c18o \ sources will have dimensions of over 0.1 pc. 
        Assuming $\theta_{source}$ $>$ 0.1\,pc and $\theta_{beam}$ = 0.124\,pc, we anticipate the beam to exceed 0.6.
        The affects of the beam filling factor on the abundance ratios and column densities are summarised in Tabel \ref{tab:factor}.
        The column density values for both  \c18o \ and \s \ are higher with the assumption of $\phi$=0.6 compared to those when $\phi$=1 is assumed.
        Nevertheless, the beam filling factor corrections  play a insignificant role in estimating $ X_{\rm ^{13}CO}$/$X_{\rm C^{18}O}$.
        
        As \co \ emission is affected by self-absorption, \c18o \ and \s \ column densities are affected by underestimates of $T_{\rm ex}$ from the \co \ data. 
        The influence of uncertainties related to the excitation temperature on the derived physical properties is investigated assuming $T_{\rm ex}(C^{18}O)$=20 K and $T_{\rm ex}(^{13}CO)$ =30\,K. This is further illustrated in Table \ref{tab:factor}.
        The column density values for both  \s \ and \c18o \ with the assumption of  $T_{\rm ex}(C^{18}O)$=20\,K and $T_{\rm ex}(^{13}CO)$ =30\,K increase compared with those where $T_{\rm ex}(^{13}CO)$ =$T_{\rm ex}(C^{18}O)$=$T_{\rm ex}(^{12}CO)$.
        However, the range of $ X_{\rm ^{13}CO}$/$X_{\rm C^{18}O}$ with $T_{\rm ex}(C^{18}O)$=20\,K and $T_{\rm ex}(^{13}CO)$ =30\,K is slightly narrower compared with those of $T_{\rm ex}(^{13}CO)$ =$T_{\rm ex}(C^{18}O)$=$T_{\rm ex}(^{12}CO)$. The higher abundance ratios ($ X_{\rm ^{13}CO}$/$X_{\rm C^{18}O}>10$) in all assumptions correspond to region D next to the W40 H\Rn2 .
        
        When including uncertainties on excitation temperatures and beam filling factors of \s \ and \c18o ,\, we are convinced that the abundance
ratio $ X_{\rm ^{13}CO}$/$X_{\rm C^{18}O}$ is lower towards more actively star-forming
regions in the AMC, with the exception of one region showing a particularly high abundance ratio (next to the W40 H\Rn2 ).

        \subsection{Comparison with H$_2$CO absorption}
        As regions B, D, and G have been identified as distinct star formation regions in the $\rm H_2CO$ absorption map of the AMC, a comparison may be made between the \c18o \  distribution and the $\rm H_2CO$ absorption \citep{2019ApJ...874..172K}. 
        Figure \ref{fig:h2co-c18o} shows a map of the $\rm C^{18}O(1-0)$ integrated intensity superposed on the integrated absorption contours of $\rm H_2CO$.
        The $\rm C^{18}O(1-0)$ and $\rm H_2CO$ data were smoothed to the same effective resolution and cell size of 10\arcmin \ and 2.5 \arcmin, respectively.
        The outer edges of the \c18o \  emission ($\thicksim$0.7\,K km $s^{-1}$) follow the lower contours of $\rm H_2CO$ ($\thicksim$0.4\,K km $s^{-1}$) and the contours encompass all identified star formation regions in the \c18o \  data.  The stronger \c18o \ emission surrounding W40 does not correspond to the $\rm H_2CO$ absorption, because the $\rm H_2CO$ absorption is related to the radio continuum of the H\,\Rn2 \ region while the enhanced \c18o \  emission results from heating by the H\,\Rn2 \ region. 
        The \c18o \  emission map at Serpens South shows several elongated structures that only partially follow the $\rm H_2CO$ absorption structure. 
        On a large scale the $\rm H_2CO$ absorption structure is similar to the $\rm C^{18}O(1-0)$ distribution but does not identify all active star formation regions.

        \section{Conclusion}
        
        Archival data of wide-field OTF observations of \co , \s , and \c18o \ (J=1-0) emission lines of the AMC were used to study star formation activity in the region and its influence on the surrounding areas.
        In the AMC, the distribution of protostellar and prestellar cores are a first indication of the locations of active star formation and their distribution is well correlated with the highest column densities within the structure of the extended  $\rm C^{18}O(1-0)$ emission.  
        This extended star formation structure is also found within the contours of the H$_2$CO absorption against the continuum of the region.

        The abundance ratio $ X_{\rm ^{13}CO}$/$X_{\rm C^{18}O}$ in our observations confirms that selective FUV photodissociation of \c18o \ indeed occurs.
        As photo-dissociation results from FUV exposure of the surrounding molecular gas, photo-dissociation is a time-dependent process that may serve as an indicator of the evolutionary age of  star formation.
        This effect is confirmed by a strong reduction in the abundance ratio $ X_{\rm ^{13}CO}$/$X_{\rm C^{18}O}$ with increasing \c18o \  column density.

        To evaluate how the beam filling factor affects the $ X_{\rm ^{13}CO}$/$X_{\rm C^{18}O}$ abundance ratio, we consider the beam filling factor to exceed 0.6 for both \c18o \ and \s \ gas.
        The abundance ratio $ X_{\rm ^{13}CO}$/$X_{\rm C^{18}O}$ distribution in the AMC was found to be similar even after the consideration of uncertainties on the beam filling factor. The beam filling factor corrections  play a insignificant role in estimating $ X_{\rm ^{13}CO}$/$X_{\rm C^{18}O}$.

        As the \co \ emission is affected by self-absorption, \c18o \ and \s \ column densities are effected by underestimates of $T_{\rm ex}$ (3.6--23.6 K) from the \co \ data.
        To study the affects of uncertainties related to excitation temperature measurements on derived physical properties,we used the assumption of $T_{\rm ex}(C^{18}O)$=20\,K and $T_{\rm ex}(^{13}CO)$ =30\,K. 
        Even if we consider higher excitation temperatures for the \s \ and \c18o \ lines,
        we come to the same conclusion, namely that with the exception of one region of  high abundance ratio (next to the W40 H\Rn2 ), the abundance ratio $ X_{\rm ^{13}CO}$/$X_{\rm C^{18}O}$ is lower toward more actively star forming regions in the AMC.

        The FUV radiation field originating at the location of massive and even fragmented star formation will also heat the surrounding molecular regions. 
        This heating process will increasingly result in 250 $\mu$m \ FIR emission and also 70 $\mu$m \ emission  after a certain evolutionary period. 
        The 70/250 $\mu$m \ flux density ratio can therefore also serve as an evolutionary age indicator and this ratio is indeed found to vary with the \c18o \  column density.

        We tested these evolutionary indicators with a number of selected star-formation and non-star-formation regions for which we estimated  the dissociation rate using the ratio $\rm X_{^{13}CO}/X_{C^{18}O}$.
        We find a clear sequence for these selected regions when correlating the 70 $\mu$m \ flux density with the estimated original mass of the regions. 
        Similarly, we find a clear sequence when correlating the 70/250 $\mu$m \ flux density ratio with the estimated photo-dissociation rate. 
        These observed parameters of the molecular medium surrounding the star formation provide a measure of the intensity of the star formation process as well as a measure of the evolutionary age of the region.
        
        \begin{acknowledgements}
                This work was sponsored by CAS-TWAS President's Fellowship for International Doctoral Students, and The National Natural Science foundation of China under grants 11433008, 11973076, 11703074, 11703073 and 11603063.
                WAB has been funded by Chinese Academy of Sciences President's International Fellowship Initiative under grants 2021VMA0008 and 2019VMA0040.
                Collaboration between ZR and the main author was supported by the University of Malaya grant UMRG FG033-017AFR.
                This publication makes use of data products from the 13.7m telescope of the Qinghai observation station of the Purple Mountain Observatory and the millimeter wave radio astronomy database,
                and the data from the Herschel Gould Belt survey (HGBS) project (http://gouldbelt-herschel.cea.fr). The HGBS is a Herschel Key Programme jointly carried out by SPIRE Specialist Astronomy Group 3 (SAG 3), scientists of several institutes in the PACS Consortium (CEA Saclay, INAF-IFSI Rome and INAF-Arcetri, KU Leuven, MPIA Heidelberg), and scientists of the Herschel Science Center (HSC).
        \end{acknowledgements}

\end{document}